\documentclass[a4paper,11pt]{article}
\pdfoutput=1 
\usepackage{jheppub}
\usepackage{epsfig}
\usepackage{amsmath}
\usepackage{graphicx}

\def\A0#1{\Pi_{\rm #1}(0)}
\def\AP0#1{\Pi'_{\rm #1}(0)}

\def\be{\begin{equation}}
\def\ee{\end{equation}}
\def\bea{\begin{array}}
\def\eea{\end{array}}
\def\beqa{\begin{eqnarray}}
\def\eeqa{\end{eqnarray}}
\def\beqas{\begin{eqnarray*}}
\def\eeqas{\end{eqnarray*}}

\def\bp{\begin{picture}}
\def\ep{\end{picture}}
\def\bc{\begin{center}}
\def\ec{\end{center}}
\def\bfig{\begin{figure}}
\def\efig{\end{figure}}

\def\bit{\begin{itemize}}
\def\eit{\end{itemize}}
\def\nn{\nonumber}
\def\f{\frac}

\def\[{\left[}
\def\]{\right]}
\def\({\left(}
\def\){\right)}

\def\..{\left.}
\def\.{\right.}
\def\tl{\tilde}
\def\ra{\rightarrow}
\def\la{\leftarrow}

\def\tm{\times}

\def\da{\dagger}

\def\la{\lambda}
\def\ta{\tau}
\def\al{\alpha}
\def\bt{\beta}

\def\ep{\epsilon}

\def\Ga{\Gamma}
\def\ga{\gamma}
\def\pa{\partial}
\def\pr{\prime}

\title{Natural solution of SUSY $\mu$ problem from modulus stabilization in modular flavor model}
\author[a]{Hong Jie Fan,}
\author[a]{Fei Wang,\footnote{Corresponding author}}
\author[a]{Ying Kai Zhang}
\affiliation[a]{School of Physics, Zhengzhou University, Zhengzhou 450000, P. R. China}
\emailAdd{feiwang@zzu.edu.cn}

\abstract{ We propose a solution to the SUSY $\mu$-problem within the framework of modular flavor symmetry. The explicit $\mu$-term is prohibited by modular symmetry, and an effective $\mu$-term is regenerated following the stabilization of the modulus field. We examine the stabilization mechanism of a single modulus field with the presence of SUSY breaking contributions
described by the non-linear SUSY realization scheme involving a nilpotent Goldstino $\textbf{X}_{nl}$ superfield.  A natural small $\mu_{eff}$, significantly smaller than the SUSY scale, can result from either the expansion of typical modular forms using a small deviation parameter near the fixed point $\omega$, or from the combined effects of suppression by powers of $q^{1/24}$ [or $(2\Im\tau)^{-1}$] along with the asymptotic suppression behavior of typical modular forms away from the fixed point $i\infty$, taking the form of appropriate power of the tiny deviation parameter. A natural small $\mu_{eff}$ can also be achieved by a weighton-like mechanism for $H_uH_d$ bilinear.  }

\begin{document}
\maketitle \indent
\newpage
\section{Introduction}
Low-energy supersymmetry (SUSY) is well motivated from a theoretical standpoint and is considered to be one of the most compelling extensions of the Standard Model. It not only address the quadratic divergence problem associated with the Higgs mass but also enables successful gauge coupling unification. Notably, the observed 125 GeV scalar, discovered by both the ATLAS~\cite{ATLAS} and CMS~\cite{CMS} collaborations of LHC, falls miraculously within the narrow $115\sim 135$ GeV 'window' predicted by the Minimal Supersymmetric Standard Model~(MSSM). This can be regarded as strong supports for low energy SUSY.

Although the framework of low energy SUSY is attractive, it still has several unsatisfactory features. In addition to the lack of conclusive signals for superparticles at the LHC, the MSSM is also bothered by the SUSY $\mu$-problem. The $\mu$-parameter, a SUSY preserving parameter in the superpotential, however, is required (without too much miraculous cancelation) to be roughly of the order of the soft SUSY breaking scale, as determined by the minimization of the MSSM Higgs potential. Many proposals have been put forward to address the $\mu$-problem, for example, by singlet extension of MSSM to Next-to-Minimal Supersymmetric Standard Model~(NMSSM) or by introducing non-renormalizable terms in the superpotential via Kim-Nilles mechanism~\cite{KN} (or in the Kahler potential via Giudice-Masiero mechanism~\cite{GM}). Most solutions involve two steps~\cite{Bae:2019dgg}: First, the appearance of $\mu$-parameter is prohibited, usually by some discrete or continuous symmetry. Second, the $\mu$ term is regenerated via some mechanism such as symmetry breaking. For detailed realizations of these two steps, one can refer to~\cite{Bae:2019dgg}.

We propose to tackle the $\mu$-problem with a similar but different approach. The symmetry adopted to forbid the $\mu$-parameter is a modular flavor symmetry. The regeneration of $\mu$ term is realized following the modular symmetry breaking, with the modulus stabilization mechanism accompanied by typical SUSY breaking contributions.

Modular symmetry is a geometrical symmetry of torus $T^2$ and it corresponds to a change of the corresponding cycle basis. Modular symmetry can naturally arise in string theory, for example, in the (exact) moduli spaces of Calabi-Yau threefolds like the quintic, and more generically in large classes of elliptic Calabi-Yau threefolds~\cite{CY:modular1,CY:modular2}.  Recently, modular flavor symmetry was proposed in~\cite{Feruglio:2017spp} to explain the flavor puzzles of the Standard Model (plus the massive neutrinos), which no longer necessitates the introduction of the flavon fields that may cause difficulties in flavor model buildings. For reviews, see~\cite{Feruglio:2019ybq,Kobayashi:2023zzc,Ding:2023htn,Ding:2024ozt}. In the modular flavor symmetry framework, the superpotential is tightly constrained by the modular symmetry, with the relevant Yukawa couplings being modular forms of a certain level $N$, which are functions of a modulus field. The matter superfields and the modular forms in the superpotential transform in some representations of certain finite modular group $\Ga_N\equiv \bar{\Ga}/\bar{\Ga}(N), N\leq 5$ (or its double cover $\Ga_N^\pr$). In this framework, the masses and mixings of quarks and leptons can be predicted with very few input parameters. Modular flavor models based on $\Ga_2\simeq S_3$~\cite{Kobayashi:2018vbk,Okada:2019xqk,Du:2020ylx}, $\Ga_3\simeq A_4$~\cite{Kobayashi:2018vbk,Petcov:2018snn,Criado:2018thu,Kobayashi:2018scp,Okada:2018yrn,
Novichkov:2018yse,Okada:2020ukr,Petcov:2022fjf,CentellesChulia:2023osj,Kumar:2023moh,Nomura:2024ctl,Pathak:2024sei}, $\Ga_4\simeq S_4$~\cite{Penedo:2018nmg,Petcov:2018snn,Novichkov:2018ovf,Kobayashi:2019xvz,Wang:2020dbp,Qu:2021jdy}, $\Ga_5\simeq A_5$~\cite{Petcov:2018snn,Novichkov:2018nkm,Ding:2019xna,Yao:2020zml,deMedeirosVarzielas:2022ihu,Abbas:2024bbv} and non-holomorphic modular flavor models~\cite{Qu:2024rns,Nomura:2024atp,Ding:2024inn,Li:2024svh} had been studied in various works to explain the low energy flavor structures of the Standard Model. Modular symmetry can also be used to solve the strong CP problem~\cite{Feruglio:2023uof,Feruglio:2024ytl}. Moreover, modular flavor symmetries can be combined with the GUT framework to further enhance the predictive power of the models, such as combining $\Ga_2,\Ga_3$ with $SU(5)$~\cite{deAnda:2018ecu,Kobayashi:2019rzp,Du:2020ylx,Chen:2021zty,Zhao:2021jxg,King:2021fhl,Ding:2021zbg}, SO(10)~\cite{Ding:2021eva,Ding:2022bzs} or flipped SU(5)~\cite{Charalampous:2021gmf,Du:2022lij,King:2024gsd} GUT models.

 The vacuum expectation value (VEV) of the modulus field can be the only source of flavour symmetry breaking. It can also be the only source of CP symmetry breaking~\cite{Baur:2019kwi,Novichkov:2019sqv,Feruglio:2021dte,Ishiguro:2020nuf,Feruglio:2024ytl}. The VEV of the modulus field $\tau$ generally break the modular symmetry completely, except at three inequivalent fixed points $\tau=i\infty, i, e^{i2\pi/3}$ in the fundamental domain. The modular forms (thus the Yukawa couplings) and the fermion mass matrices are fixed once the modular flavour symmetry is fully or partially broken by the VEV of $\tau$.
From a top-down perspective, the value of modulus VEV can be determined dynamically with a modular-invariant scalar potential in 4d $N=1$ supergravity. This is also crucial for giving nonzero masses to the modulus field, rendering it phenomenological acceptable.

Experiences from heterotic string theory on moduli stabilization~\cite{target:space,Leedom:2022zdm,Gonzalo:2018guu} indicate that the scalar potential favors minima at the boundary of the fundamental domain or at points inside the fundamental domain that are close to the fixed point $\omega=e^{i 2\pi/3}$ with negative vacuum energy. In the presence of additional "matter"-like fields, it is possible to obtain a dS vacua with additional uplifting contributions, such as the introduction of an anti-D3 brane that provides a positive uplifting potential, which would induce
a SUSY breaking vacuum shift to a tiny positive vacuum energy. The presence of the anti-brane-like sources can be described in terms of a nilpotent Goldstino superfield~\cite{KS} in a supersymmetric effective field theory. In the framework of non-linearly realized SUSY, the couplings of chiral, gauge and gravity superfields to the constrained nilpotent Goldstino superfield $\textbf{X}_{nl}$ can be used to describe spontaneous SUSY breaking in
supergravity~\cite{Burgess:2021juk,Aparicio:2015psl,NL:SUGRA1,NL:SUGRA2,NL:SUGRA3,NL:SUGRA4,NL:SUGRA5,NL:SUGRA6} and the presence of the anti-brane type sources with a supersymmetric action. Contributions from nilpotent chiral superfield can uplift the scalar potential responsible for the stabilization of the  modulus field.

This paper is organized as follows. In Sec~\ref{sec-2}, we briefly review the $\Ga_N$ modular flavor symmetry. In Sec~\ref{sec-3}, we discuss the solution of SUSY $\mu$-problem using modular flavor symmetry. In Sec~\ref{sec-4}, we discuss the stabilization of singlet modulus field in the presence of SUSY breaking contributions described by the non-linear realized SUSY scheme with the nilpotent Goldstino ${\textbf{X}}_{nl}$ superfield. In Sec~\ref{sec-5}, we discuss the regeneration of a naturally small $\mu_{eff}$ after modulus stabilization, when the modulus field is stabilized near some fixed points, such as $\omega\equiv e^{i2\pi/3}$ and $i\infty$ . Sec~\ref{sec:conclusions} contains our conclusions.
\section{\label{sec-2} Modular $\Ga_N$ symmetry}

The full modular group is defined as the set of linear fractional transformations acting on the complex modulus $\tau$ in the upper half complex plane
\begin{equation}
\ga=\(\begin{array}{cc}
a  &  b  \\
c  &  d
\end{array}\):\tau\longmapsto\gamma\tau=\frac{a\tau+b}{c\tau+d},~~\text{with}~~a, b, c, d\in\mathbb{Z},~~ad-bc=1,~~~{\rm Im}\tau>0\,.
\end{equation}
 This group can be generated by two generators, $S$ and $T$, which act as follows
\begin{equation}
S:\tau\mapsto -\frac{1}{\tau},~~~~\quad T: \tau\mapsto\tau+1\,.
\end{equation}
These generators satisfy the relations
\beqa
S^4=1,~~(ST)^3=1,~~S^2T=TS^2.
\eeqa
The $S,T$ generators can be represented by the following $SL(2,Z)$ matrices,
\begin{equation}
S=\left(\begin{array}{cc}
0 & 1 \\
-1  & 0
\end{array}
\right),~~~\quad T=\left(\begin{array}{cc}
1 & 1 \\
0  & 1
\end{array}
\right)~.
\end{equation}
 Finite modular groups $\Ga_N$~\cite{Feruglio:2017spp} are defined as $\Ga_N=\overline{\Ga}/\overline{\Gamma}(N)$, where $\overline{\Gamma}\equiv SL(2,Z)/\{\pm 1\}$ and
$\overline{\Gamma}(N)$ is given by
\beqa
\overline{\Gamma}(N)=\left\{\bea{c}\Gamma(N),~~~~~~~~~~~N\geq3\\ \Gamma(N)/\{\pm 1\},~~~ N=2\eea\right.~~~,
\eeqa
with $\Gamma(N)$ the principal congruence subgroup of level $N$
\begin{equation}
\Gamma(N)=\left\{\begin{pmatrix}
a  & b \\
c  & d
\end{pmatrix}\in SL(2, \mathbb{Z}),~ b=c=0\,(\mathrm{mod~N}), a=d=1\,(\mathrm{mod~N})
\right\}\,.
\end{equation}
Finite modular groups $\Ga_N$ can be obtained by further imposing the condition $T^{N}=1$ in addition to $S^2=(ST)^3=1$.

  Modular form $f(\tau)$ of weight $2k$ and level $N$ is a holomorphic function of the complex modulus $\tau$ with well-defined transformation properties under the action of ${\Gamma}(N)$
\begin{equation}
f\left(\frac{a\tau+b}{c\tau+d}\right)=(c\tau+d)^{2k} f(\tau)~, ~~~\forall~\ga\equiv\left(
\begin{array}{cc}
a  &  b  \\
c  &  d
\end{array}
\right)\in{\Gamma}(N)\,.
\end{equation}

Modular forms of weight $2k$ and level $N$ span a linear space of finite dimension. They can be arranged into some modular multiplets $Y_{\mathbf{r}}^{(2k)}$ that transform as irreducible representation $\mathbf{r}$ of the finite modular group $\Gamma_N$ up to automorphy factor~\cite{Feruglio:2017spp}, i.e.
\begin{equation}
Y_{\mathbf{r}}^{(2k)}(\gamma\tau)=(c\tau+d)^{2k}\rho_{\mathbf{r}}(\gamma)Y_{\mathbf{r}}^{(2k)},~~~~\forall~\gamma\in\overline{\Gamma}\,,
\end{equation}
where $\gamma$ is the representative element of the coset
$\gamma \overline{\Gamma}(N)$ in $\Gamma_N$, and $\rho_{\mathbf{r}}(\gamma)$ is the representation matrix of the element $\gamma$ in the irreducible representation $\mathbf{r}$.
The actions of modular group on supermultiplets read
\beqa
\phi_i(\ga\tau)=(c\tau+d)^{-k_\phi} \rho^{\mathbf{r^\pr}}_{ij}(\gamma)\phi_j~,~~~\forall \ga\in \bar{\Ga}~,
\eeqa
where $k_\phi$ denotes the modular weight of $\phi$, and $\rho$ is a unitary representation of the quotient group $\bar{\Ga}_N$.

\section{\label{sec-3} Natural $\mu$-term in modular flavor model}
As noted in~\cite{Bae:2019dgg}, to solve the SUSY $\mu$ problem, the appearance of a $\mu$ term in the superpotential should be prohibited by some symmetry. We propose that the superfields $H_u$, $H_d$ transform non-trivially under modular transformations, thereby prohibiting the ordinary $\mu H_u H_d$ term by modular symmetry.

The relevant Lagrangian can be given as
\beqa
{\cal L}&\supseteq& \int d^4\theta \[-3\log(2 \Im \tau)+ \f{H_u^\da H_u}{(2 \Im\tau)^{k_u}}
+ \f{H_d^\da H_d}{(2 \Im\tau)^{k_d}}+\cdots\]\nn\\
&+&\int d^2\theta \[\al \Lambda_{SUSY} \f{Y_N^{(k_r)}(\tau)}{\[\eta(\tau)\]^{2k_3}} H_u H_d+\cdots\],
\label{mu:superpotential}
\eeqa
where the coefficient for bilinear $H_uH_d$ term is given by
\beqa
\tl{\mu}=\al \Lambda_{SUSY} \f{Y_N^{(k_r)}(\tau)}{\[\eta(\tau)\]^{2k_3}}~,
\eeqa
for $k_{H_u}+k_{H_d}=k_r-k_3+3$ to ensure the modular invariance of
\beqa
e^G\equiv e^K |W|^2~.
\eeqa
Here, $\al\sim {\cal O}(1)$ is a dimensionless parameter and $\Lambda_{SUSY}$ is typically of order the SUSY scale. The Dedekind $\eta$-function is a modular form with modular weight $1/2$ under modular transformation up to a phase. Its transformation under $\ga$ is given by
 \beqa
\eta(\ga\tau)=\epsilon_{\eta}(\ga)(c\tau+d)^{1/2} \eta(\tau)~, ~~\forall \ga\in \overline{\Ga}~,
\eeqa
where $\epsilon_{\eta}(\ga)$ depends on $\ga$ (but not on $\tau$) and can be expressed as
\beqa
\epsilon_{\eta}(\ga)&=&\exp\[\f{\pi i}{12}\omega(a,b,c,d)-i\f{\pi}{4}\],~
\eeqa
that satisfies $\epsilon_{\eta}(\ga)^{24}=1$~\cite{DHoker:2022dxx}. Here
\beqa
\omega(a,b,c,d)=\(\f{a+d}{c}+12s(-d,c)\),
\eeqa
is an integer and $s(-d,c)$ is a Dedekind sum.

The expansion of the $\eta(\tau)$ function is
\beqa
\eta(\tau)&=&q^{1/24}\sum\limits_{n=-\infty}^{\infty}(-1)^n q^{n(3n-1)/2},\nn\\
&=&q^{1/24}\[1+\sum\limits_{n=1}^{\infty}(-1)^n\(q^{n(3n-1)/2}+q^{n(3n+1)/2}\)\],\nn\\
&=&q^{1/24}\[1-q-q^2+q^5+q^7-q^{12}-\cdots\],~~~~~q\equiv e^{i 2\pi \tau}~.
\eeqa

It is evident that the Kahler potential for superfields with non-trivial modular weights, such as $H_u$ and $H_d$ superfields, is non-minimal. Therefore, after the modulus field $\tau$ acquires a VEV through the modulus stabilization mechanism, the $H_u, H_d$ fields need to be normalized properly to get their canonical kinetic forms. This normalization corresponds to the following replacements
\beqa
H_u\ra H_u (2 \Im\tau)^{k_{H_u}/2}~,~~~H_d\ra H_d (2 \Im\tau)^{k_{H_d}/2}~.
\eeqa
Thus, when the modulus field is stabilized, the effective $\mu$-parameter is regenerated as
\beqa
\mu_{eff}=\al \Lambda_{SUSY} \f{Y_N^{(k_r)}(\tau_0)}{\[\eta(\tau_0)\]^{2k_3}} (2 \Im\tau_0)^{(k_{H_u}+k_{H_d})/2}=\al \Lambda_{SUSY} \f{Y_N^{(k_r)}(\tau_0)}{\[\eta(\tau_0)\]^{2k_3}} (2 \Im\tau_0)^{(k_r-k_3+3)/2}.
\eeqa
We need to explain why $\mu_{eff}$ is much smaller than $\Lambda_{SUSY}$ and of order the electroweak scale or soft SUSY breaking scale $m_{soft}$. We will show that, the smallness of $\mu_{eff}$ is the consequence of large suppression factors, either originate from some power of deviation parameter that characterizes the deviation of modulus VEV from typical fixed point, or from  non-trivial power (determined by the choice of modular weights for $H_u$ and $H_d$) of a small factor related to the modulus VEV. All these suppression factors are determined by the modulus stabilization mechanism, with the stabilized modulus VEVs being affected by SUSY breaking effects. The smallness of $\mu_{eff}$ can also be achieved by a weighton-like mechanism~\cite{King:2020qaj} for $H_u H_d$ bilinear.

\section{\label{sec-4} Stabilization of modulus field with nilpotent Goldstino ${\textbf{X}}_{nl}$ superfield}
To stabilize the modulus field, we can utilize effective supergravity scalar potentials derived from a weight $-3$ superpotential to determine the minimum. The modulus VEV can always be restricted to the fundamental domain
\begin{equation}
\label{eq:fund_domain}
{D}_H \equiv \left\{ \tau \in \mathcal{H} : -\frac{1}{2} \leq \Re {\tau} < \frac{1}{2},\, |\tau| > 1 \right\} \cup \left\{ \tau \in \mathcal{H} : -\frac{1}{2} < \Re \tau \leq 0,\, |\tau| = 1 \right\} \,,
\end{equation}
of the modular group, as any choice of the modulus VEV in the upper-half plane can be uniquely related to a $\tau\in D_H$ in the fundamental domain via a modular transformation, which is physically equivalent to such a choice. Discussions on modulus stabilization with single modulus and multiple moduli can be found in~\cite{Kobayashi:2019xvz,Kobayashi:2020hoc,Kobayashi:2020uaj,Ishiguro:2020tmo,NPP:2201.02020,PLNSR:2304.14437,King:2023snq,King:2024ssx,Ding:2024neh,Higaki:2024pql,Higaki:2024jdk,Kobayashi:2023spx,Funakoshi:2024yxg}.

To ensure the modular invariance and Kahler transformation invariance of
\beqa
e^G\equiv e^K |W|^2~,
\eeqa
the superpotential for the modulus should take the form
\beqa
W(\tau)=c_1^3\f{H(\tau)}{\[\eta(\tau)\]^{6}},
\eeqa
with the Kahler potential
\beqa
K_{mod}=-3\log(2\Im\tau)~,
\eeqa
in the unit $M_{Pl}=1$. Here, $c_1M_{Pl}$ denotes the typical mass scale for the superpotential, where $c_1$ is a dimensionless number.

 The modular invariant holomorphic function $H(\tau)$ is given by
\beqa
 H(\tau)_{m,n}&=&\(j(\tau)-1728\)^{m/2}j(\tau)^{n/3}{\cal P}\(j(\tau)\),\nn\\
 &\equiv&\(\f{G_6(\tau)}{\[\eta(\tau)\]^{12}}\)^m\(\f{G_4(\tau)}{\[\eta(\tau)\]^{8}}\)^n{\cal P}\(j(\tau)\)~,
 \label{modular:invariant}
\eeqa
where ${\cal P}\(j(\tau)\)$ denotes a polynomial with respect to $j(\tau)$ and $m,n$ are non-negative integers~\cite{target:space}. The definitions of $G_4(\tau)$ and $G_6(\tau)$ are given in the Appendix~\ref{appendix:B}. The Klein $j$-function is defined as a modular form of zero weight
\beqa
j(\tau)&\equiv& \f{3^65^3}{\pi^{12}}\f{G_4(\tau)^3}{\[\eta(\tau)\]^{24}},\nn\\
&\approx& 744+q^{-1}+196884 q+ 21493760 q^2 + 864299970 q^3 + {\cal O}(q^4)~,
\eeqa
with $j(i\infty)$ diverges, $j(\omega) = 0$ and $j(i) = 1728$. Note that $j(\tau)$ has a triple zero at $\tau= \omega$, and $j(\tau)-1728$ has a double zero at $\tau=i$. The derivative $j^\pr(\tau)$ vanishes only at these values: $j^\pr(\omega) = j^\prime(i) = 0$~\cite{Klein:jfunction}.

We should note that, in such a stabilization mechanism, the fixed points $z=i, \omega\equiv e^{i2\pi/3}$ are always the extrema of the scalar potential, although they may not be the global minimum.
With proper SUSY breaking contributions, possibly from the "matter"-like sector, the stabilized modulus value, which located at/near the fixed points $z_{FP}=\omega, i$, can slightly deviate from the original global minimum. The minimum of the scalar potential can also possibly be uplifted to dS~(de Sitter) type. In fact, the classical 10d supergravity Maldacena-Nunez no-go result~\cite{MN} suggests that any constructions leading to a dS vacuum must include non-classical ingredients, such as D-branes, O-planes or quantum corrections. To describe the scenario in which SUSY is linearly realized in a gravitational sector and non-linearly realized in the uplifting SUSY breaking sector (for example, from anti-D branes), the formalism of non-linearly realized SUSY as formulated by~\cite{KS} can be used for such a system. It was proved in~\cite{Burgess:2021juk} that, with the non-linear SUSY realization scheme, the supergravity form is stable even if the supermultiplets in SUSY breaking sector are split badly enough to allow some of its members to be integrated out while the others are not.

When SUSY is spontaneously broken in supergravity, the Goldstino is eaten by the gravitino via the super-Higgs mechanism. When such a breaking happens at low energies
compared with the Planck scale, the Goldstino couplings can be conveniently described by a (constrained) independent superfield in the supergravity effective action. The Goldstino provides a way for an arbitrary non-supersymmetric low-energy theory to realize SUSY non-linearly.
 The emergence of the constraints may also be viewed as an effective field theory limit in which certain mass parameters become very large and the corresponding modes decouple.

The coupling of non-linearly realized SUSY with a nilpotent chiral Goldstino superfield, subjected to constraints ${\textbf{X}}_{nl}^2=0$, can be described by
\beqa
{\cal L}\supseteq \int d^4\theta \textbf{X}_{nl}^\da \textbf{X}_{nl}+\[\int d^2\theta f_0 \textbf{X}_{nl}+h.c.\]~.
\eeqa
The nilpotent constraints for the chiral Goldstino superfield can be solved to be
\beqa
\textbf{X}_{nl}=\f{\psi_X\psi_X}{2 F_X}+\sqrt{2} \theta \psi_X+\theta^2 F_X~,
\eeqa
where all the fields are the functions of $y^\mu\equiv x^\mu+i\theta\sigma^\mu\bar{\theta}$. The lowest component of $\textbf{X}_{nl}$ is eliminated by the nilpotent constraints $\textbf{X}_{nl}^2=0$ and is no longer an independent complex scalar field. This leaves only the Goldstino field $\psi_X$ and an independent auxiliary field $F_X$ in $\textbf{X}_{nl}$ superfield. The Goldstino field $\psi_X$, which is the fermionic component of $\textbf{X}_{nl}$, eventually mixes with the gravitino to give it a mass through the super-Higgs mechanism. The lowest component of $\textbf{X}_{nl}$, which is denoted as ${X}_{nl}$ for the Goldstino bilinear ${\psi_X\psi_X}$, obviously satisfies ${X}_{nl}^2=0$ due to its anti-commuting Grassman variable nature. The chiral superfield $\textbf{X}_{nl}$ can be seen as the long-distance expression of the microscopic superfield $\textbf{X}$ that satisfies $\bar{D}^{\dot{\al}}{\cal J}_{\al\dot{\al}}=D_\al \textbf{X}$ for Ferrara-Zumino multiplet ${\cal J}_{\al\dot{\al}}$~\cite{KS}, which can be used even when SUSY is spontaneously broken and can be followed consistently along the renormalization group flow down to the deep IR. As noted in~\cite{Aparicio:2015psl}, for the anti-D3 brane
in the KKLT~\cite{Kachru:2003aw} scenario, the representation in terms of $\textbf{X}_{nl}$ is very convenient and enough to capture all the anti-D3 brane degrees of freedoms since it allows to treat the effect of anti-D3 brane in terms of standard supergravity couplings of matter and moduli superfields to the nilpotent Goldstino superfield. From the perspective of low energy effective field theory, regardless of the UV origin of SUSY breaking, such a non-linear realization scheme is also convenient for discussions on modulus stabilization and vacuum uplifting.

 When coupled to supergravity, the most general form (at the
two-derivative level) for the action in superspace is given by
\beqa
{\cal L}\supseteq \int d^2\Theta 2 {\cal{E}}\left[\f{3}{8}\( \overline{\cal{D}}^2 - 8 {\cal R} \) \, e^{-K/3} + W \right] + \hbox{h.c.} \,,
\eeqa
with
\beqa
K=\textbf{X}_{nl}^\da \textbf{X}_{nl}~,~~~W=f_0 \textbf{X}_{nl}+c_1^3~,
\eeqa
where without loss of generality we take $f_0$ and $c_1$ to be real. Here $\cal{E}$ is the chiral density, ${\cal R}$ is the the superspace curvature and $\overline{\cal{D}}^2 - 8 {\cal R}$ is the covariant generalization of the chiral projection operators $D^\ga D_\ga$ in supergravity~\cite{Wess:Bagger}.

Integrating out the auxillary field $F_X$ gives $F_X =-(W_{\textbf{X}_{nl}})^* = -f_0$. To accommodate the modular flavor symmetry, the Kahler potential and superpotential needs to be adapted to
\beqa
K&\supseteq& -3\log(2\Im\tau)+\f{\textbf{X}^\da_{nl}\textbf{X}_{nl}}{(2\Im \tau)^{k_X}}~,\nn\\
W&\supseteq& \(f_0 \textbf{X}_{nl} \[\eta(\tau)\]^{2k_X}+c_1^3\)\f{H(\tau)}{[\eta(\tau)]^6}.
\label{Kah:Super}
\eeqa
To ensure the modular invariance of the superpotential, the modular weight $k_X$ should satisfy $k_X\in 12\mathbb{Z}$.
The supergravity scalar potential can be derived from
\beqa
V=e^K\[K^{\al\bar{\bt}}\nabla_\al W \nabla_{\bar{\bt}} W- 3W \overline{W}\],
\label{SUGRA:potential}
\eeqa
where the Kahler-covariant derivative of the superpotential is
\beqa
\nabla_\al W=W_\al+K_\al W~.
\eeqa
From the Kahler potential, we can calculate the inverse Kahler metric
\beqa
K^{{\al}\bar{\beta}}
&=&\(\bea{cc}\f{(2\Im\tau)^{2}}{3}\[1-\f{k_X}{3}\f{{X}^*_{nl}{X}_{nl}}{(2\Im\tau)^{k_X}}\]&\f{i(2\Im\tau) k_X {X}_{nl}}{3}\\
-\f{i(2\Im\tau) k_X {X}^*_{nl}}{3}&(2\Im\tau)^{k_X}\[1+\f{ k_X^2 ({X}^*_{nl}{X}_{nl})}{3(2\Im\tau)^{k_X}}\]
\eea\)~.
\eeqa
Here the indices $\al,\bt\sim (\tau,{X}_{nl})$ and the nilpotency of ${X}_{nl}$ is used in the calculation.

From the general expression of the SUGRA scalar potential in (\ref{SUGRA:potential}), the scalar potential for the modulus field $\tau$ and $X_{nl}$ can be calculated as
\beqa
&&V(X_{nl},{X}^*_{nl},\tau,{\tau}^*)\nn\\
&=&\f{e^{\f{|X_{nl}|^2}{(2\Im\tau)^{k_X}}}}{(2\Im\tau)^3|\eta(\tau)|^{12}}\left\{\f{(2\Im \tau)^2}{3}\(1-\f{k_X}{3}\f{|X_{nl}|^2}{(2\Im\tau)^{k_X}}\)\left|\tl{W}_M \tl{c}_M+H(\tau)\tl{W}_T\right|^2\right.\nn\\
&+& (2\Im\tau)^{k_X}\[1+\f{ k_X^2 |X_{nl}|^2}{3(2\Im\tau)^{k_X-1}}\]\left|H(\tau)\tl{W}_X\right|^{2}\nn\\
&+& 2\Re\[\f{i(2\Im\tau) k_X X_{nl}}{3}(\tl{W}_X H(\tau))^*\(\tl{W}_M \tl{c}_M+H(\tau)\tl{W}_T\)\]\nn\\
&-&\left.3\left|\tl{c}_M\right|^2\left|{H(\tau)}\right|^2\right\}~,
\label{general:scalarpotential}
\eeqa
where
\beqa
\tl{W}_M &\equiv& H^\pr(\tau)-\f{3 i H(\tau)}{2\pi}\hat{G}_2(\tau,\tau^*)~,\nn\\
\tl{W}_T &\equiv& k_X f_0 X_{nl}\[\eta(\tau)\]^{2k_X}\[\f{i}{2\pi}\hat{G}_2(\tau,\tau^*)+\f{i}{2\Im \tau}+\f{i c_1^3X_{nl}^*}{f_0(2\Im \tau)^{k_X+1}\[\eta(\tau)\]^{2k_X}}\]~,\nn\\
\tl{c}_{M}&\equiv&f_0 X_{nl} \[\eta(\tau)\]^{2k_X}+c_1^3~,\nn\\
\tl{W}_X&\equiv&f_0\[\eta(\tau)\]^{2k_X}\left(1+\f{|X_{nl}|^2}{(2\Im\tau)^{k_X}}\right)+\f{{X}^*_{nl}c_1^3}{(2\Im\tau)^{k_X}}.
\label{general:definitions}
\eeqa
The constraint $X_{nl}^2=0$ for the lowest component of $\textbf{X}_{nl}$ is used to eliminate some of the high order $X_{nl}$ terms. We also use the following expression for the derivative of Dedekind $\eta$ function
 \beqa
 \f{\eta^\pr(\tau)}{\eta(\tau)}=\f{i}{4\pi} G_2(\tau)~,
 \eeqa
 where the holomorphic Eisenstein series $G_2$, which is a quasi-modular form
 \beqa
 G_2(\ga \tau)=(c\tau+d)^2G_2(\tau)-2\pi i c(c\tau+d)~,
 \eeqa
 is related to its weight two non-holomorphic counterpart $\hat{G}_2(\tau,\bar{\tau})$ as
\beqa
\hat{G}_2(\tau,{\tau}^*)=G_2(\tau) -\f{\pi}{\Im \tau}~.
\eeqa
From the definition of $H(\tau)$, its derivative with respect to $\tau$ can be calculated to be
\beqa
\label{Htau:derivative}
H'(\tau) &=& \left\{j(\tau)^{n/3}[j(\tau)-1728]^{m/2} \mathcal{P}(j(\tau))\right\}^\prime~, \nn\\
&=&\left\{\f{n}{3}j(\tau)^{n/3-1}[j(\tau)-1728]^{m/2} \mathcal{P}(j(\tau))+\f{m}{2}j(\tau)^{n/3}[j(\tau)-1728]^{m/2-1}\mathcal{P}(j(\tau))\right.\nn\\
&+& \left.j(\tau)^{n/3}[j(\tau)-1728]^{m/2} \mathcal{P}^\pr(j(\tau))\right\}j^\pr(\tau)~.
\label{Hprime:expression}
\eeqa
For non-vanishing values of $H(\tau)$, we have
\beqa
H'(\tau)&=& H(\tau) j'(\tau) \left[ \frac{n}{3} \frac{1}{j(\tau)} + \frac{m}{2} \frac{1}{j(\tau)-1728} + \frac{1}{\mathcal{P}(j(\tau))} \frac{\partial \mathcal{P}(j(\tau))}{\partial j(\tau)} \right],
\eeqa
making obvious that $H(\tau)$ can be factored out from $\tl{W}_M$ as
\beqa
\label{WM:factorization}
\tl{W}_M&=&H(\tau)\left( j'(\tau) \left[ \frac{n}{3} \frac{1}{j(\tau)} + \frac{m}{2} \frac{1}{j(\tau)-1728} + \frac{1}{\mathcal{P}(j(\tau))} \frac{\partial \mathcal{P}(j(\tau))}{\partial j(\tau)} \right] -\f{3 i }{2\pi}\hat{G}_2(\tau,\tau^*)\right).\nn
\eeqa
Substituting the factorized form of $\tl{W}_M$ back into the scalar potential, it becomes apparent that the $|H(\tau)|^2$ factor can be extracted from the expression when $H(\tau)\neq 0$. This observation plays an important role in locating the extrema of the scalar potential.

For $m=n=0$, the trivial case
\beqa
H'(\tau) &=& \mathcal{P}^\pr(j(\tau))j^\pr(\tau)~,
\eeqa
vanishes at $z=\omega,i$ because $j^\pr(\omega)=j^\pr(i)=0$.

For $m=1$ or $n=1,2$, from the results
\beqa
\f{j^\pr(i)}{[j(i)-1728]^{1/2}}&=&-i 2\pi  \f{[E_4(i)]^2}{[\eta(i)]^{12}},~~\f{j^\pr(\omega)}{[j(\omega)]^{1/3}}=-i 2\pi \f{E_4(\omega)E_6(\omega)}{[\eta(\omega)]^{16}}=0~,\nn\\
~\f{j^\pr(\omega)}{[j(\omega)]^{2/3}}&=&-i 2\pi \f{E_6(\omega)}{[\eta(\omega)]^{8}}~,
\label{Hprime:iomega}
\eeqa
we can see that $H^\pr(i)$ and $H^\pr(\tau)$ are finite with such choices of $m,n$,
within which we use
\begin{equation}
\frac{\partial}{\partial \tau} j(\tau)= -i 2\pi \frac{E_6(\tau)}{E_4(\tau)}j(\tau) = -i 2\pi \frac{E_6(\tau)[E_4(\tau)]^2}{[\eta(\tau)]^{24}} ,
\end{equation}
and the relations~\cite{target:space}
\beqa
j(\tau)-1728=\(\f{E_6(\tau)}{[\eta(\tau)]^{12}}\)^2~,~~~j(\tau)=\(\f{E_4(\tau)}{[\eta(\tau)]^{8}}\)^3~,
\eeqa
as well as the values of $\eta$-function at $z=i,\omega$
\beqa
\eta(i)=\f{\Ga(\f{1}{4})}{2\pi^{3/4}}~,~~~~\eta(\omega)=e^{-i\f{\pi}{24}}\frac{3^{\f{1}{8}} [\Gamma(\f{1}{3})]^{\f{3}{2}}}{2 \pi}~,
\eeqa
and some special value for $E_4(\tau)$ and $E_6(\tau)$~\cite{DHoker:2022dxx}
\beqa
E_4(i)=\f{48[\Ga(\f{5}{4})]^4}{\pi^2[\Ga(\f{3}{4})]^4},~~
E_4(\omega)= 0,~~E_6(\omega)=\f{729}{2\pi^3}\f{[\Ga(\f{4}{3})]^6}{[\Ga(\f{5}{6})]^6}.
\eeqa

For $m>1$ and $n>2$, it is obvious that $H^\pr(\tau)$ is not singular at zero points of $H(\tau)$, therefore vanishes at $i$ and $\omega$, as $j^\pr(i)=j^\pr(\omega)=0$.

So, we have
\beqa
H^\pr(i)&=&\left\{\bea{c}-i\pi  (12)^{n}\f{[E_4(i)]^2}{[\eta(i)]^{12}}P(1728) , ~~~m=1\\ 0,~~~~~~~~~~~~~~~~\quad~\quad\quad\quad\quad~~m\neq 1\eea\right.\nn\\
H^\pr(\omega)&=&\left\{\bea{c}-\f{i 2\pi}{3} (-12)^{3m/2}\f{E_6(\omega)}{[\eta(\omega)]^{8}}P(0),~~~n=1\\0 ,~~~~~~~~~~~~~~~~~\quad\quad\quad\quad\quad~~n\neq 1\eea\right.~.
\label{Hprime}
\eeqa


The condition $H(\tau)= 0$ can be reduced to
\begin{equation}
j(\tau)^{n/3}=0,\quad\text{or}\quad [j(\tau)-1728]^{m/2} = 0, \quad \text{or} \quad \mathcal{P}(j(\tau)) = 0.
\end{equation}
The corresponding solutions are $\tau=i,\omega$ (for $m,n\neq 0$) and $\tau=\tilde{\tau}_i$ that satisfies $P(j(\tilde{\tau}_i))=0$. Whereas $H'(\tau)$ typically vanishes at $\tau = i$ and $\tau = \omega$ (barring special choices of $m$ and $n$), it generally does not vanish at $\tau = \tilde{\tau}_i$, the point where $P(j(\tilde{\tau}_i)) = 0$. However, for some special choices of $P(j(\tilde{\tau}))$, $H^\pr(\tau)$ can still vanish at $\tau=\tilde{\tau}_0$. For example, when
\beqa
P(j(\tau))=\prod\limits_{i}(j(\tau)-j(\tilde{\tau}_i))^2~,
\eeqa
then $\tau=\tilde{\tau}_0$ acts as a multiple root of $P(j(\tilde{\tau}))=0$ and $H^\pr(\tau)$ vanishes at such roots. We will discuss such interesting possibilities in the next subsection.

The generic, complicated form of the scalar potential may obscure the discussions regarding the stabilization mechanism. Therefore, we would like to concentrate on its typical simplification forms.

\subsection{The case with $k_X=0$}

When $k_X=0$, the scalar potential can be simplified significantly. It is evident that, due to the presence of the $K_\alpha W$ terms within the Kahler-covariant derivative of the superpotential $\nabla_\alpha W$, the scalar potential still includes non-trivial interference terms between the $X_{nl}$ sector and the modulus field $\tau$ sector, despite the fact that the Kahler metric $K^{\alpha\bar{\beta}}$ is diagonal in this case.

The potential can then be written as
  \beqa
  &&V(X_{nl},{X}^*_{nl},\tau,{\tau}^*)\nn\\
&=&\f{e^{|X_{nl}|^2}}{(2\Im\tau)^3|\eta(\tau)|^{12}}\left\{\f{(2\Im \tau)^2}{3}\left|\tl{W}_M\tl{c}_M\right|^2\right.\nn\\
&+&\left.\left|H(\tau)\right|^{2}
\left|f_0\left(1+{|X_{nl}|^2}\right)+{{X}^*_{nl}c_1^3}\right|^2  -3\left|\tl{c}_M\right|^2\left|{H(\tau)}\right|^2\right\}~,\nn\\
&=&\f{1}{(2\Im\tau)^3|\eta(\tau)|^{12}}\left\{\f{(2\Im \tau)^2}{3}\left|\tl{W}_M\right|^2\(f_0^2|X_{nl}|^2+c_1^6+2\Re\[f_0 X_{nl}c_1^3\]\)\right.\nn\\
&+&\left|H(\tau)\right|^{2}
\( f^2_0+2f_0^2{|X_{nl}|^2}+2\Re\[f_0c_1^3{X}_{nl}^*\] +|X_{nl}|^2c_1^6\f{}{}\)~\nn\\  &-&\left.3\left|{H(\tau)}\right|^2\(f_0^2|X_{nl}|^2+c_1^6+2\Re\[f_0 X_{nl}c_1^3\]\)\right\}~\nn\\
&+&\f{|X_{nl}|^2}{(2\Im\tau)^3|\eta(\tau)|^{12}}\left\{\f{(2\Im \tau)^2}{3}\left|\tl{W}_M\right|^2c_1^6+\left|H(\tau)\right|^{2}f^2_0-3c_1^6\left|{H(\tau)}\right|^2\right\}~,
 \label{potential:k=0}
  \eeqa
where we adopt
\beqa
\tl{W}_M&=& H^\pr(\tau)-\f{3 i H(\tau)}{2\pi}\hat{G}_2(\tau,\tau^*)~,\nn\\
\tl{c}_{M}&=&f_0 X_{nl}+c_1^3~,
\eeqa
from the definitions in (\ref{general:definitions}).

In this scenario, the modulus VEV can be determined by the extreme conditions of the scalar potential
\beqa
\f{\pa }{\pa\tau}V(X_{nl},{X}^*_{nl},\tau,{\tau}^*)=\f{\pa }{\pa\tau^*}V(X_{nl},{X}^*_{nl},\tau,{\tau}^*)=0~,\label{V:tau}\\
\f{\pa }{\pa X_{nl}}V(X_{nl},{X}^*_{nl},\tau,{\tau}^*)=\f{\pa }{\pa X_{nl}^*}V(X_{nl},{X}^*_{nl},\tau,{\tau}^*)=0~.\label{V:Xnl}
\eeqa
 Given that the scalar potential is real, the two equations on the right follow directly from those on the left via complex conjugation. To determine if a local extremum is a minimum, we need to assess the positive definiteness of the $4\times 4$ Hessian matrix
\beqa
(H_f)_{ij}\equiv \f{\pa^2 V}{\pa x_i\pa x_j}~,~~~{\rm for}~~x_i={\Re}\tau, {\Im}\tau,{\Re}X_{nl},{\Im}X_{nl}~,
\eeqa
at that extremum. The elements of the Hessian matrix can be rewritten by
\beqa
\f{\pa^2 }{\pa x_1^2}V &=&2\(\f{\pa^2}{\pa\tau\pa\tau^*}+\Re\f{\pa^2 }{\pa \tau\pa\tau}\)V(X_{nl},{X}^*_{nl},\tau,{\tau}^*),\nn\\
\f{\pa^2 }{\pa x_2^2}V &=&2\(\f{\pa^2}{\pa\tau\pa\tau^*}-\Re\f{\pa^2 }{\pa \tau\pa\tau}\)V(X_{nl},{X}^*_{nl},\tau,{\tau}^*),\nn\\
\f{\pa^2 }{\pa x_1\pa x_2}V &=&-2\Im \f{\pa^2 }{\pa \tau\pa\tau}V(X_{nl},{X}^*_{nl},\tau,{\tau}^*),
\eeqa
with similar terms for $X_{nl}$ and $X_{nl}^*$ in terms of $\rho$ and $\theta$
\beqa
\frac{\partial^2}{\partial x_3^2} V&=&2\(\f{\pa^2}{\pa X_{nl}\pa X_{nl}^*}+\Re\f{\pa^2 }{\pa X_{nl}\pa X_{nl}}\)V(X_{nl},{X}^*_{nl},\tau,{\tau}^*),\\
 &=& \(\cos^2 \theta \frac{\partial^2}{\partial \rho^2} - 2 \frac{\cos \theta \sin \theta}{\rho} \frac{\partial^2}{\partial \rho \partial \theta} + \frac{\sin^2 \theta}{\rho^2} \frac{\partial^2}{\partial \theta^2} + \frac{\sin^2 \theta}{\rho} \frac{\partial}{\partial \rho} + 2 \frac{\cos \theta \sin \theta}{\rho^2} \frac{\partial}{\partial \theta}\) V~,\nn\\
\frac{\partial^2}{\partial x_4^2} V &=& 2\(\f{\pa^2}{\pa X_{nl}\pa X_{nl}^*}-\Re\f{\pa^2 }{\pa X_{nl}\pa X_{nl}}\)V(X_{nl},{X}^*_{nl},\tau,{\tau}^*),\nn\\
&=&\( \sin^2 \theta \frac{\partial^2}{\partial \rho^2} + 2 \frac{\sin \theta \cos \theta}{\rho} \frac{\partial^2}{\partial \rho \partial \theta} + \frac{\cos^2 \theta}{\rho^2} \frac{\partial^2}{\partial \theta^2} + \frac{\cos^2 \theta}{\rho} \frac{\partial}{\partial \rho} - 2 \frac{\sin \theta \cos \theta}{\rho^2} \frac{\partial}{\partial \theta}\) V~,\nn\\
\frac{\partial^2}{\partial x_3 \partial x_4} V &=&-2\Im \f{\pa^2 }{\pa X_{nl}\pa X_{nl}}V(X_{nl},{X}^*_{nl},\tau,{\tau}^*),\nn\\
&=&\( \sin \theta \cos \theta \frac{\partial^2}{\partial \rho^2} + \frac{\cos 2\theta}{\rho} \frac{\partial^2}{\partial \rho \partial \theta} - \frac{\sin \theta \cos \theta}{\rho^2} \frac{\partial^2}{\partial \theta^2} - \frac{\sin \theta \cos \theta}{\rho} \frac{\partial}{\partial \rho} - \frac{\cos 2\theta}{\rho^2} \frac{\partial}{\partial \theta}\) V~.\nn
\eeqa
Other second derivatives of $V(X_{nl},{X}^*_{nl},\tau,{\tau}^*)$ with respect to $\tau,\tau^*$ and $\rho, \theta$ are given by
\beqa
\f{\pa^2 }{\pa x_1\pa x_3}V
&=&\(\f{\pa^2}{\pa \tau \pa X_{nl}}+\f{\pa^2}{\pa \tau^* \pa X_{nl}^*}+\f{\pa^2}{\pa \tau \pa X_{nl}^*}+\f{\pa^2}{\pa \tau^* \pa X_{nl}}\)V(X_{nl},{X}^*_{nl},\tau,{\tau}^*)~,\nn\\
&=&\[\cos\theta\(\f{\pa^2}{\pa \tau \pa \rho}+\f{\pa^2}{\pa \tau^* \pa \rho}\)-\f{\sin\theta}{\rho}\(\f{\pa^2}{\pa \tau \pa \theta}+\f{\pa^2}{\pa \tau^* \pa \theta}\)\]V(X_{nl},{X}^*_{nl},\tau,{\tau}^*)~,\nn\\
\f{\pa^2 }{\pa x_2\pa x_4}V
&=&-\(\f{\pa^2}{\pa \tau \pa X_{nl}}+\f{\pa^2}{\pa \tau^* \pa X_{nl}^*}-\f{\pa^2}{\pa \tau \pa X_{nl}^*}-\f{\pa^2}{\pa \tau^* \pa X_{nl}}\)V(X_{nl},{X}^*_{nl},\tau,{\tau}^*)~,\nn\\
&=&-\[i\sin\theta\(\f{\pa^2}{\pa \tau \pa \rho}-\f{\pa^2}{\pa \tau^* \pa \rho}\)+i\f{\cos\theta}{\rho}\(\f{\pa^2}{\pa \tau \pa \theta}-\f{\pa^2}{\pa \tau^* \pa \theta}\)\]V(X_{nl},{X}^*_{nl},\tau,{\tau}^*)~,\nn\\
\f{\pa^2 }{\pa x_1\pa x_4}V
&=&-i\(\f{\pa^2}{\pa \tau \pa X_{nl}}-\f{\pa^2}{\pa \tau^* \pa X_{nl}^*}+\f{\pa^2}{\pa \tau \pa X_{nl}^*}-\f{\pa^2}{\pa \tau^* \pa X_{nl}}\)V(X_{nl},{X}^*_{nl},\tau,{\tau}^*)~,\nn\\
&=&\[\sin\theta\(\f{\pa^2}{\pa \tau \pa \rho}+\f{\pa^2}{\pa \tau^* \pa \rho}\)+\f{\cos\theta}{\rho}\(\f{\pa^2}{\pa \tau \pa \theta}+\f{\pa^2}{\pa \tau^* \pa \theta}\)\]V(X_{nl},{X}^*_{nl},\tau,{\tau}^*)~,\nn\\
\f{\pa^2 }{\pa x_2\pa x_3}V
&=&i\(\f{\pa^2}{\pa \tau \pa X_{nl}}-\f{\pa^2}{\pa \tau^* \pa X_{nl}^*}-\f{\pa^2}{\pa \tau \pa X_{nl}^*}+\f{\pa^2}{\pa \tau^* \pa X_{nl}}\)V(X_{nl},{X}^*_{nl},\tau,{\tau}^*)~,\\
&=&\[i\cos\theta\(\f{\pa^2}{\pa \tau \pa \rho}-\f{\pa^2}{\pa \tau^* \pa \rho}\)-i\f{\sin\theta}{\rho}\(\f{\pa^2}{\pa \tau \pa \theta}-\f{\pa^2}{\pa \tau^* \pa \theta}\)\]V(X_{nl},{X}^*_{nl},\tau,{\tau}^*)~.\nn
\eeqa

The analytical expression for each element of the Hessian matrix is highly complex and only takes a simplified form at specific points, such as the fixed points $\omega$ and $i$. Since this study does not focus on the fixed points, we evaluate the positivity of the eigenvalues for the Hessian matrix at the extrema numerically, rather than through analysis of its analytical form, in the subsequent discussions.

The extreme condition (\ref{V:tau}) $V_\tau=0$ can be calculated as
\beqa
\f{\pa}{\pa \tau} V
&=&\f{\f{3i}{2\pi}\hat{G}_2(\tau,\tau^*)}{(2\Im\tau)^3|\eta(\tau)|^{12}}\left\{\f{(2\Im \tau)^2}{3}\left|\tl{W}_M\tl{c}_M\right|^2\right.\nn\\
&+&\left.\left|H(\tau)\right|^{2}
\left|f_0\left(1+{|X_{nl}|^2}\right)+{{X}^*_{nl}c_1^3}\right|^2  -3\left|\tl{c}_M\right|^2\left|{H(\tau)}\right|^2\right\}\nn\\
&+&\f{1}{(2\Im\tau)^3|\eta(\tau)|^{12}}
\left\{\f{-4i\Im\tau}{3}\left|\tl{W}_M\tl{c}_M\right|^2+\f{(2\Im \tau)^2}{3}\left|\tl{c}_M\right|^2\tl{W}_M^*\tl{W}_M^\pr\right.\nn\\
&+&\left. H(\tau)^*H^\pr(\tau)\(\left|f_0\left(1+{|X_{nl}|^2}\right)+{{X}^*_{nl}c_1^3}\right|^2 -3\left|\tl{c}_M\right|^2\)
\right\}\nn\\
&+&\f{\f{3i}{2\pi}\hat{G}_2(\tau,\tau^*)|X_{nl}|^2}{(2\Im\tau)^3|\eta(\tau)|^{12}}\left\{\f{(2\Im \tau)^2}{3}\left|\tl{W}_M\tl{c}_M\right|^2+\left|H(\tau)\right|^{2}\(f_0^2 -3\left|\tl{c}_M\right|^2\)\right\}\nn\\
&+&\f{{|X_{nl}|^2}}{(2\Im\tau)^3|\eta(\tau)|^{12}}
\left\{\f{-4i\Im\tau}{3}\left|\tl{W}_M\tl{c}_M\right|^2+\f{(2\Im \tau)^2}{3}\left|\tl{c}_M\right|^2\tl{W}_M^*\tl{W}_M^\pr\right.\nn\\
&+&\left. H(\tau)^*H^\pr(\tau) \(f_0^2-3\left|\tl{c}_M\right|^2\)
\right\},
\label{extrema:tau}
\eeqa
and its complex conjugate for $V_{\tau^*}$.

Within this expression, $\tl{W}_M^\pr$ is given by
\beqa
\tl{W}_M^\pr&=& H^{\pr\pr}(\tau)-\f{3 i H^\pr(\tau)}{2\pi}\hat{G}_2(\tau,\tau^*)
-\f{3 i H(\tau)}{2\pi}\hat{G}^\pr_2(\tau,\tau^*)~.
\eeqa

When the value of $\tau$ satisfies $H(\tau)\neq 0$, $\tl{W}_M^\pr$ can be rewritten as
\beqa
\tl{W}_M^\pr&=&H(\tau) [j'(\tau)]^2 \left[ \frac{n}{3} \frac{1}{j(\tau)} + \frac{m}{2} \frac{1}{j(\tau)-1728} + \frac{1}{\mathcal{P}(j(\tau))} \frac{\partial \mathcal{P}(j(\tau))}{\partial j(\tau)} \right]^2\nn\\
&+&H(\tau)\left\{j^{\pr}(\tau)\left[ \frac{n}{3} \frac{1}{j(\tau)} + \frac{m}{2} \frac{1}{j(\tau)-1728} + \frac{1}{\mathcal{P}(j(\tau))} \frac{\partial \mathcal{P}(j(\tau))}{\partial j(\tau)} \right]\right\}^\pr~\nn\\
&-&H(\tau)\f{3 i }{2\pi}\hat{G}_2(\tau,\tau^*)[j'(\tau)]^2 \left[ \frac{n}{3} \frac{1}{j(\tau)} + \frac{m}{2} \frac{1}{j(\tau)-1728} + \frac{1}{\mathcal{P}(j(\tau))} \frac{\partial \mathcal{P}(j(\tau))}{\partial j(\tau)} \right]\nn\\
&-&\f{3 i H(\tau)}{2\pi}\hat{G}^\pr_2(\tau,\tau^*)~,
\label{WM:derivative}
\eeqa
from (\ref{Htau:derivative}). The derivative of $\hat{G}^\pr_2(\tau,\tau^*)$ with respect to $\tau$ is given as follows
\beqa
\hat{G}^\pr_2(\tau,\tau^*)=\f{i\pi^3}{18}\(E_2(\tau)^2- E_4(\tau)\)-i\f{2\pi}{(2\Im\tau)^2}~.
\eeqa
 From equations (\ref{WM:factorization}) and (\ref{WM:derivative}), one can check that $|H(\tau)|^2$ acts as a positive multiplicative factor in $V_\tau$ when $H(\tau)\neq 0$.
In addition to the solutions of $V_\ta/|H(\tau)|^2=0$ (for $H(\tau)\neq 0$), the constraints $\tl{W}_M^\pr=0$ can be satisfied also by some solutions of $H(\tau)=0$, which we need to check point by point.

From equation (\ref{Hprime:expression}), we have the following discussions on the condition $H^\pr(\tau)=0$ when $H(\tau)= 0$:
\bit
\item The case $\tau=i$ is a solution of $H(\tau)= 0$:

When $\tau=i$ is a solution of $H(\tau)= 0$ for $m>0$, the condition $H'(\tau) = 0$ can be satisfied automatically for $m\neq 1$, see equation (\ref{Hprime}).
 It can also be satisfied when $\mathcal{P}(1728)=0$ for $m=1$, requiring $P(j(\tau))$ to take the form
 \beqa
 P(j(\tau))\propto (j(\tau)-1728)^{k}~, ~~~~k\in \mathbb{N}~.
 \label{PjForm:m}
 \eeqa
 Such factor of $P(j(\tau))$ can be absorbed into $(j(\tau)-1728)^{m/2}$ term in $H(\tau)$ after the redefinition of '$m$', rendering the redefined $\mathcal{P}(1728)\neq 0$. So, we conclude that $H'(i) = 0$ can be satisfied for $m> 1$. Note that $H'(i) = 0$ for $m=0$, although $\tau=i$ is in general not a solution of $H(\tau)= 0$ when $m=0$.
\item The case $\tau=\omega$ is a solution of $H(\tau)= 0$:

Similarly, when $\tau=\omega$ is a solution of $H(\tau)= 0$ for $n>0$, the condition $H'(\tau) = 0$ can be satisfied automatically for $n\neq 1$. It can also be satisfied when $\mathcal{P}(0)=0$ for $n=1$, which requires $P(j(\tau))$ to take the form
 \beqa
 P(j(\tau))\propto  [j(\tau)]^{k}~, ~~~~k\in \mathbb{N}~.
 \label{PjForm:n}
 \eeqa
After the redefinition of '$n$', we conclude that the condition $H'(\omega) = 0$ can be satisfied for $n> 1$. Note also $H'(\omega) = 0$ when $n=0$, although $\tau=\omega$ is in general not a solution of $H(\tau)= 0$ when $n=0$..

\item The case $\tau = \tilde{\tau}_i$ with $P(j(\tilde{\tau}_i)) = 0$:

The condition $H'(\tau) = 0$ can be satisfied for $P^\pr(j(\tilde{\tau}_i))=0$, which requires $\tau=\tilde{\tau}_i$ being a root of $P(j(\tau))$ with
multiplicity two or greater. Therefore,
\beqa
 P(j(\tau))\propto  \pm g(j(\tau)) \prod\limits_{i} (j(\tau)-j(\tilde{\tau}_i))^{k_i}~,~~~~k_i\in \mathbb{N}~.
\label{PjForm:first}
 \eeqa
with $k_i \geq 2$ and $g(j(\tau))$ an arbitrary positive polynomial function of $j(\tau)$.

\eit

Before we discuss the condition $H^{\pr\pr}(\tau)=0$ when $H(\tau)=0$, we need to know the values of $H^{\pr\pr}(\tau)$ at the fixed points. It can be calculated that
 \beqa
 H^{\pr\pr}(i)&=&\left\{\bea{c}-\f{2\pi^2}{[\eta(i)]^{24}}[E_4(i)]^4\[\f{n}{3}P(1728)(12)^{n-3}+(12)^{n} P^\pr(1728)\]~,~~~~m=0\\
 ~0,~~~~~~~~~~~~~~~~~~~~~~~~~~~~~~~~~~~~~~~~~~~~~~~~~~~~~~~~~~~~~~~~~~~~~~~m=1\\
 -\f{2\pi^2}{[\eta(i)]^{24}}[E_4(i)]^4 (12)^n P(1728)~,~~~~~~~~~~~~~~~~~~~~~~~~~~~~~~~~~~~~~m=2\\
 ~0,~~~~~~~~~~~~~~~~~~~~~~~~~~~~~~~~~~~~~~~~~~~~~~~~~~~~~~~~~~~~~~~~~~~~~~~~~~m>2\eea\right.\nn\\
 H^{\pr\pr}(\omega)&=&\left\{\bea{c}0~,~~~~~~~~~~~~~~~~~~~~~~~~~~~~~~~~~~~~~~~~~~~~~~~~~~~~~~~~~~~~~n=0,1,3,4,5 \\
 -\f{8\pi^2}{9}\f{[E_6(\omega)]^2}{[\eta(\omega)]^{16}}(-12)^{\f{m}{2}}P(0)~,~~~~~~~~~~~~~~~~~~~~~~~~~~~~~~~~~~~~~~~~~n=2\\
 0~.~~~~~~~~~~~~~~~~~~~~~~~~~~~~~~~~~~~~~~~~~~~~~~~~~~~~~~~~~~~~~~~~~~~~~~~n>5
 \eea\right.
 \label{Hprimeprime}
 \eeqa
 Here we use the following relations
 \beqa
 \f{j^{\pr\pr}(\omega)}{[j(\omega)]^{\f{1}{3}}}&=&-\f{8\pi^2}{3}\f{[E_6(\omega)]^2}{[\eta(\omega)]^{16}}~,
 ~~~\f{j^{\pr\pr}(\omega)}{[j(\omega)]^{\f{2}{3}}}=-\f{8\pi^2}{3[\eta(\omega)]^{8}}\f{[E_6(\omega)]^2}{E_4(\omega)}\ra \infty~,~\nn\\
 \f{j^{\pr\pr}(i)}{[j(i)-1728]^{\f{1}{2}}}&=&-\f{2\pi^2}{[\eta(\omega)]^{12}}\f{[E_4(i)]^4}{E_6(i)} \ra \infty~,
 \eeqa
 and
\beqa
\f{[j^\pr(i)]^2}{[j(i)-1728]^{2-\f{m}{2}}}&=&-\f{4\pi^2}{[\eta(\omega)]^{12m}}\f{[E_6(i)]^{2}[E_4(i)]^4}{[E_6(i)]^{4-m}},~~~\nn\\
\f{[j^\pr(\omega)]^2}{[j(\omega)]^{2-\f{n}{3}}}&=&-\f{4\pi^2}{[\eta(\omega)]^{8n}}\f{[E_6(\omega)]^2[E_4(\omega)]^4}{[E_4(\omega)]^{6-n}}~.
\eeqa
to cancel each divergent terms among the $j^{\pr\pr}(\tau)$ and $(j^\pr(\tau))^2$ related terms within $H^{\pr\pr}(\tau)$ at the fixed points when $m=1$ or $n=1$.

When $H(\tau)= 0$, we have the following discussions on the condition $H^{\pr\pr}(\tau)=0$:
\bit
\item The case $\tau=i$ is a solution of $H(\tau)= 0$:

When $\tau=i$ is a solution of $H(\tau)= 0$ for $m>0$, the condition $H^{\pr\pr}(\tau) = 0$ can be satisfied automatically for $m\neq 2$.
For other choices of $m$, it can also be satisfied when $\mathcal{P}(1728)=\mathcal{P}^\pr(1728)=0$, requiring $P(j(\tau))$ to take the form
 \beqa
 P(j(\tau))\propto (j(\tau)-1728)^{k+1}~, ~~~~k\in \mathbb{N}~.
 \eeqa
 After the redefinition of $m$, we conclude that $H^{\pr\pr}(i) = 0$ can be satisfied for $m\neq 0,2$.
\item The case $\tau=\omega$ is a solution of $H(\tau)= 0$:

Similarly, when $\tau=\omega$ is a solution of $H(\tau)= 0$ for $n>0$, the condition $H^{\pr\pr}(\tau) = 0$ can be satisfied automatically for $n\neq 2$. For other choices of $m$, it can also be satisfied when $\mathcal{P}(0)=0$, requiring $P(j(\tau))$ to take the form
 \beqa
 P(j(\tau))\propto  [j(\tau)]^{k}~, ~~~~k\in \mathbb{N}~.
 \eeqa
 After the redefinition of $n$, we conclude that $H^{\pr\pr}(\omega) = 0$ can be satisfied for $n\neq 0,2$.

\item The case $\tau = \tilde{\tau}_i$ with $P(j(\tilde{\tau}_i)) = 0$:

The condition $H^{\pr\pr}(\tau) = 0$ can be satisfied for $P^\pr(j(\tilde{\tau}_i))=P^{\pr\pr}(j(\tilde{\tau}_i))=0$, which requires $\tau=\tilde{\tau}_i$ being a root of $P(j(\tau))$ with
multiplicity three or greater. Therefore, the most general form of $P(j(\tau))$ should be
 \beqa
 P(j(\tau))=\pm g(j(\tau)) \prod\limits_{i} (j(\tau)-j(\tilde{\tau}_i))^{k_i}~,~~~~k_i\in \mathbb{N}~.
 \label{PjForm}
 \eeqa
with $k_i \geq 3$ and $g(j(\tau))$ an arbitrary positive polynomial function of $j(\tau)$.
\eit

It is obvious from (\ref{extrema:tau}) that the stationary condition for $\tau$ is determined by the solutions of $W_M^*W_M^\pr=0$ and $|W_M|^2=0$ when $H(\tau)=0$. From (\ref{Hprime}) and (\ref{Hprimeprime}), it is obvious that $W_M^*W_M^\pr=0$ always vanish at $\tau=i$ except when $m=1$ (or $\tau=\omega$ except when $n=1$). However, they will not vanish at $\tau=\tilde{\tau}_i$ in general, where $P(j(\tilde{\tau}_i))=0$. Therefore, solutions of equation $V_\tau/|H(\tau)|^2=0$ (for $H(\tau)\neq 0$) and fixed points  $\tau=i,\omega$ are always stationary points with respect to $\tau$ and $\tau^*$. As the solutions of combined equations $V_\tau/|H(\tau)|^2=0$ (for $H(\tau)\neq 0$) and (\ref{V:Xnl}) in general can not be expressed in analytical forms and need numerical calculations, it is hard to get the desired location of the modulus VEV for our model buildings. So, we would like to concentrate on other solutions that can be determined analytically, with the corresponding modulus VEVs deviating slightly from the fixed points. Such non-fixed point solutions of modulus VEVs can come from $H(\tau)= 0$.

The value $\tau = \tilde{\tau}_i$ with $P(j(\tilde{\tau}_i)) = 0$ generally fails to be a solution of $V_\tau = 0$ because the derivatives $H'(\tau)$ (or the expression $[H^{\pr\pr}(\tau)-\f{3 i H^\pr(\tau)}{2\pi}\hat{G}_2(\tau,\tau^*)]$) are typically non-zero at this point. In order for $H^\pr(\tilde{\tau}_i)=0$, $P(j(\tau))$ needs to take the form (\ref{PjForm:first}). In such a case, non-fixed point values of $\tau$ with $P(j(\tau))=0$ can also satisfy $P^{\pr}(j(\tau))=0$ to guarantee $\tl{W}_M^\pr\tl{W}^*_M=0$. We will concentrate on such interesting choices of $P(j(\tau))$ in our subsequent studies unless otherwise specified. As the $ H(\tau)$ (with $m,n\neq 1$) factor can be effectively factored out from $\tl{W}_M^\pr$ and $\tl{W}_M$, any non-fixed point value of $\tau$ satisfying $|H(\tau)|^2=0$ is an extremum of the potential, as long as the condition $V_{X_{nl}}=0$ is met.

It was argued in~\cite{target:space,Ding:2024neh,Higaki:2024pql} that modular invariance and $\tau\leftrightarrow-\tau^*$ symmetry of the scalar potential ensure that the first derivative of scalar potential on $\tau$ along certain directions at the boundary of the fundamental domain vanishes
\beqa
\left.\f{\pa }{\pa x_1} V\right|_{x_1=\pm1/2,0}=0~,~~~\left.\f{\pa }{\pa \rho_\tau} V\right|_{\rho_\tau=1}=0~,~~~
\label{extrema}
\eeqa
where $\tau\equiv x_1+i x_2\equiv \rho_\tau e^{i\theta_\tau}$. As the two independent directional derivatives vanishes at the finite fixed points $\tau=\omega$ and $\tau=i$, the scalar potential is always stationary with respect to $\tau$ at such finite fixed points~\cite{King:2023snq,Higaki:2024pql}, agreeing with our previous discussions.

The extrema conditions (\ref{V:Xnl}) can be rewritten as
\beqa
\f{\pa}{\pa \rho} V=\f{\pa}{\pa \theta}V=0~,
\eeqa
with $X_{nl}\equiv\rho e^{i\theta}$. We have the following discussions for some special parameter choices:
\bit

\item {\bf Case} $f_0=0$:

The extrema of the scalar potential are located at parameter points where $X_{nl} = 0$. The modulus VEV associated with these extrema is identical to the case involving only a single modulus field $\tau$, which had been extensively studied in previous works~\cite{target:space,NPP:2201.02020,Leedom:2022zdm}. It has been demonstrated that the minima of the scalar potential can slightly deviate from the left (right) cusp symmetric point and the boundary when $m \neq 0$ and $n = 0$ for $H(\tau)$, resulting in negative vacuum energy.

\item {\bf Case} $c_1=0$:

The extrema of the scalar potential are also located at the parameter points with $\rho=0$ (that is, $X_{nl} = 0$). The corresponding potential energy is given by
\beqa
\left.V\right|_{X_{nl}=0}=\f{1}{(2\Im\tau)^3|\eta(\tau)|^{12}}\left|{H(\tau)}\right|^2 f_0^2\geq 0~,
\eeqa
for all values of $\tau$ within the fundamental region.  The parameter points for local minimum of the scalar potential should satisfy both $H(\tau) = 0$ and $X_{nl}=0$. As $|H(\tau)|^2$ can be effectively factored out from $V_\tau$ with $P(j(\tau))$ taking the form~(\ref{PjForm}), solutions of $H(\tau) = 0$ indeed satisfy $V_\tau = 0$ in general (except for typical choices of $m,n\leq 2$ at the fixed points). Therefore, in this case, the stabilized modulus VEVs are located at the solutions of $H(\tau) = 0$, corresponding to vanishing vacuum energy.

\item {\bf Case} $f_0\neq 0$, $c_1\neq 0$:

We now focus on the general case with $f_0, c_1 \neq 0$. It can be calculated that the extremum condition for $\theta$ is
\begin{equation}
\frac{\partial}{\partial \theta} V = -\frac{2 f_0 c_1^3}{(2\Im\tau)^3|\eta(\tau)|^{12}} \rho \sin\theta
\left[\frac{(2\Im \tau)^2}{3}\left|\widetilde{W}_M\right|^2 - 2\left|H(\tau)\right|^{2}\right] = 0~,
\label{extreme:theta}
\end{equation}
while the extremum condition for $\rho$ is
\beqa
\frac{\partial}{\partial \rho} V &=& \frac{1}{(2\Im\tau)^3|\eta(\tau)|^{12}}\left\{
2\rho\[\frac{(2\Im \tau)^2}{3}\left|\widetilde{W}_M\right|^2\(f_0^2+c_1^6\)
-2c_1^6\left|H(\tau)\right|^{2}\] \right.\nn\\
&+&\left. 2f_0c_1^3\cos\theta\[\frac{(2\Im \tau)^2}{3}\left|\widetilde{W}_M\right|^2-2\left|H(\tau)\right|^{2}\]\right\}
= 0~.
\label{extreme:rho}
\eeqa
It should be noted that the point $\tau = i\infty$ does not satisfy the extremum condition for $\theta$ since $[(2\Im\tau)|\eta(\tau)|^{4}]^{-1}$ diverges as $\Im\tau \to \infty$.


As discussed previously, the $|H(\tau)|^2$ factor can be extracted from $\left|\widetilde{W}_M\right|^2$ for non-vanishing $H(\tau)$. Besides, $\left|\widetilde{W}_M\right|^2$ vanishes when $H(\tau)= 0$ for choices of $P(j(\tau))$ in the form~(\ref{PjForm}) and $m,n\neq 1$. Therefore,  $|H(\tau)|^2=0$, which in general satisfies the extremum condition $V_\tau=0$ for $\tau$, can also in general satisfy the extremum conditions $V_\rho=V_\theta=0$ for $X_{nl}$. So we can conclude that modulus VEVs that satisfy $|H(\tau)|^2=0$ for any values of $X_{nl}$ are always local extrema of the scalar potential.

The extremum conditions for $\theta$ and $\rho$ can also be satisfied in the following cases:
\begin{itemize}
     \item From (\ref{extreme:theta}), for $\sin\theta\neq 0$ (that corresponds to a non-real $X_{nl}$ with $\theta\neq 0,\pi$), we can get the condition for modulus $\tau$
    \begin{equation}
    \frac{(2\Im \tau)^2}{3}\left|\widetilde{W}_M\right|^2 - 2\left|H(\tau)\right|^{2} = 0,
    \end{equation}
    which is independent of the $\rho$ value. After substituting this condition into (\ref{extreme:rho}), we can obtain the extremum condition for the $\rho$ parameter:
    \begin{equation}
    \frac{\partial}{\partial \rho} V = 4\rho f_0^2 \frac{\left|H(\tau)\right|^{2}}{(2\Im\tau)^3|\eta(\tau)|^{12}} = 0.
    \end{equation}
    Such an extremum condition for $\rho$ can be satisfied with
    \begin{equation}
    \left|H(\tau)\right|^{2} = 0.
    \end{equation}

Thus, we have the extremum conditions for the case with $f_0, c_1 \neq 0$ and a non-vanishing non-real $X_{nl}$ value:
\begin{equation}
\left|\widetilde{W}_M\right|^2 = \left|H(\tau)\right|^{2} = 0.
\label{extreme:rho-theta}
\end{equation}
Substituting back into the scalar potential, we can see that the potential energy at this extremum vanishes.


Then, we would like to focus on the constraint
\begin{equation}
\widetilde{W}_M \equiv H'(\tau) - \frac{3 i H(\tau)}{2\pi} \hat{G}_2(\tau, \tau^*) = 0,
\end{equation}
which satisfies $H(\tau) = 0$. Discussions on the solutions of $H'(\tau)=0$ are given in Appendix \ref{appendix:C}. For the fixed points $\tau = \omega, i$, they are always the local extrema of the scalar potential when $m, n>1$. When $m<1$ or $n<1$,  whether the fixed points $\tau = \omega, i$ are local extrema can easily be figured out case by case, with the relevant results agreeing with the single modulus case in~\cite{Leedom:2022zdm}. The extrema conditions (\ref{extreme:rho-theta}) and (\ref{extrema:tau}) of the scalar potential can also be satisfied for non-fixed points solutions of $H(\tau) = 0$ when $P(j(\tau))$ takes the form~(\ref{PjForm:first}), and the corresponding vacua are always of Minkowski type. The Minkowski vacua determined by $H(\tau) = 0$ are independent of the specific value of $X_{nl}$. To identify the global minimum of the scalar potential, it is necessary to evaluate the potential energies for all extrema, including the case where $X_{nl}$ is real.

 \item Real values of $X_{nl}$ with $\theta= 0$ or $\theta=\pi$.

In this case, the scalar potential reduces to
\begin{eqnarray}
&&V(X_{nl}, X^*_{nl}, \tau, \tau^*) \nonumber \\
&=& \frac{1}{(2\Im\tau)^3 |\eta(\tau)|^{12}} \left\{ \frac{(2\Im \tau)^2}{3} \left| \tilde{W}_M \right|^2 \left( f_0^2 \tilde{X}^2 + c_1^6 + 2f_0 \tilde{X} c_1^3 \right) \right. \nonumber \\
&+& \left| H(\tau) \right|^{2} \left( f^2_0 + 2f_0^2 \tilde{X}^2 + 2 f_0 c_1^3 \tilde{X} + \tilde{X}^2 c_1^6 \right) \nonumber \\
&-& \left. 3 \left| H(\tau) \right|^2 \left( f_0^2 \tilde{X}^2 + c_1^6 + 2 f_0 \tilde{X} c_1^3 \right) \right\} \nonumber \\
&+& \frac{\tilde{X}^2}{(2\Im\tau)^3 |\eta(\tau)|^{12}} \left\{ \frac{(2\Im \tau)^2}{3} \left| \tilde{W}_M \right|^2 c_1^6 + \left| H(\tau) \right|^{2} f^2_0 - 3 c_1^6 \left| H(\tau) \right|^2 \right\},
\end{eqnarray}
for real values of $X_{nl}$, which we denote simply as $\tl{X}$. It is a quadratic polynomial of $\tl{X}$, which can be rewritten as
\beqa
 V(\tl{X}, \tau, \tau^*)=\f{1}{(2\Im\tau)^3|\eta(\tau)|^{12}}\[C_1\tl{X}^2+ C_2\tl{X}+C_3\]~,
 \label{quadratic:potential}
 \eeqa
 where
 \beqa
 C_1&=&\f{(2\Im \tau)^2}{3}\left|\tl{W}_M\right|^2(f_0^2+c_1^6)-2\left|H(\tau)\right|^{2} c_1^6~,\nn\\
 C_2&=&\f{(2\Im \tau)^2}{3}\left|\tl{W}_M\right|^2 2 f_0c_1^3-4\left|H(\tau)\right|^{2}f_0c_1^3~,\nn\\
 C_3&=&\f{(2\Im \tau)^2}{3}\left|\tl{W}_M\right|^2 c_1^6+\left|H(\tau)\right|^{2}\(f_0^2-3c_1^6\)~.
 \eeqa

Obviously, the scalar potential is real for all (real) values of $\widetilde{X}$.
The scalar potential is bounded from below for $\widetilde{X} \in \mathbb{R}$ when the coefficients of the quadratic polynomial in~(\ref{quadratic:potential}) satisfy $C_1 > 0$ (and $C_2 = 0$ when $C_1 = 0$, rendering the scalar potential finite).

For $C_1 > 0$, the scalar potential has a minimum with the corresponding $\widetilde{X}$ value and the potential energy
\beqa
\label{minimum:realX}
\left.V\right|_{min;\tl{X}=\f{-C_2}{2C_1}}&=& C_3-\f{C_2^2}{4C_1}~.~
\eeqa
To determine the extrema of $\left.V\right|_{min}$, we need to determine
\beqa
\f{\pa}{\pa \tau} \[C_3-\f{C_2^2}{4C_1}\]=\f{\pa}{\pa \tau^*}\[C_3-\f{C_2^2}{4C_1}\]=0~.
\eeqa
These extremum conditions, after factoring out a common non-vanishing $|H(\tau)|^2$
term, yield non-trivial equations for $\tau$ that generally require numerical solution.

For $C_1 = 0$, it can be checked that $C_2 = 0$ when stationary condition $H(\tau)=0$ for modulus VEV $\tau$ holds, which corresponds to Minkowski vacua for all values of $\tl{X}$. Therefore, in line with earlier discussions, VEVs of modulus field that satisfy $H(\tau)=0$ for all values of
 ${X}_{nl}$ (whether real or complex) represent local minima of the scalar potential.

  The pseudo-modulus $X_{nl}$, i.e. a continuous space of  vacua
with degenerate tree-level vacuum energies, can be lifted by SUSY breaking effects. Taking into account soft SUSY breaking mass term $m_{X}^2 |X_{nl}|^2$ from additional SUSY breaking sources with proper mediation mechanism (or from the one-loop Coleman-Weinberg potential of $X_{nl}$ after including its couplings to matter fields, for example, of the O'Raifeartaigh type $[\la_{ij} X_{nl}+M_{ij}]\phi_i\phi_j$ in the superpotential), the scalar potential for $X_{nl}$ taking the form $V(X_{nl})\approx |\tilde{C}|^2 f_0^2+ m_{X}^2|X_{nl}|^2$ can stabilize the pseudo-modulus at $X_{nl}=0$ in the flat SUSY limit, with $\tilde{C}$ some constant determined by the stabilized value of $\tau$.

For $C_1 < 0$, the scalar potential, being a quadratic polynomial in $\widetilde{X}$, is not bounded from below. From a physical perspective, the bounded-from-below requirement $C_1 > 0$ can be relaxed to allow $C_1 \leq 0$ because of the constraint $|\widetilde{X}|\leq M_{\text{Pl}}$. In such cases, the minimum of the scalar potential at fixed $\tau$ is located at $\widetilde{X} \sim M_{\text{Pl}}$ (or $\widetilde{X} \sim -M_{\text{Pl}}$) when $C_1 \leq 0$. Higher-order and quantum gravity effects can substantially modify the shape of the scalar potential for field values of $|\widetilde{X}|$ close to $M_{\text{Pl}}$, rendering $\widetilde{X} \sim \pm M_{\text{Pl}}$ local minima of the scalar potential. To determine the global minimum of the scalar potential, one must compare the potential energies for real $X_{nl}$ with those of the Minkowski vacua determined by the condition $H(\tau) = 0$.


\end{itemize}

For the case $C_1 > 0$, the minimum (\ref{minimum:realX}) is always located at some value of $\widetilde{X}$ with $\widetilde{X}\ll M_{Pl}$. This minimum can either be the global minimum or merely a local minimum when the bounded-from-below condition is relaxed by allowing $C_1 \leq 0$.

\begin{figure}[hbt]
\begin{center}
\includegraphics[width=2.9 in]{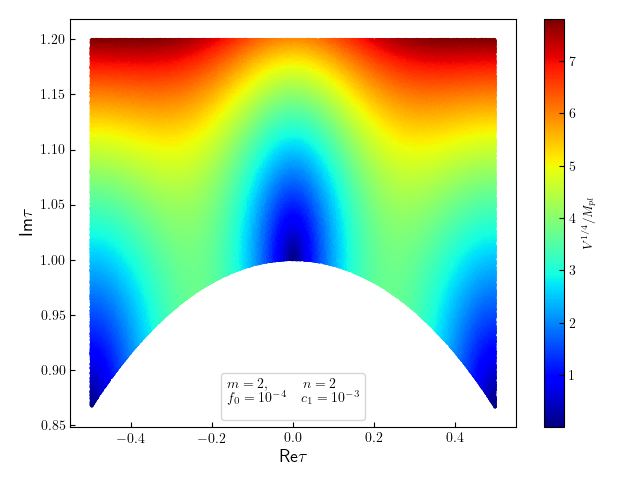}
\includegraphics[width=2.9 in]{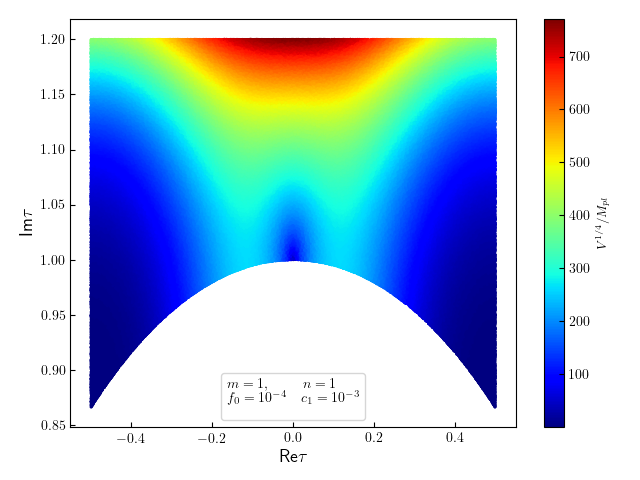}\\
\vspace{-.5cm}\end{center}
\caption{The values of $V(X_{nl},\bar{X}_{nl},\tau,{\tau}^*)$ within the fundamental domain in the scenario with $k_X=0$ and $X_{nl}=0$ for various parameter choices when $f_0\neq 0$ and $f_0\gg c_1^2$. We adopt $\mathcal P(j(\tau))=1$ for the left panel and $\mathcal P(j(\tau))=(j(\tau)-j(\tau_0))^2$ with $\tau_0=e^{i\f{2\pi}{3}(1-0.006)}$ for the right panel.}
\label{fig1}
\end{figure}

 We select some benchmark parameter points with various choices of $f_0,c_1$ and calculate numerically the potential energies for all extrema to identify the global minimum. For many benchmark points, the global minimum is determined by the condition $H(\tau) = 0$ with vanishing vacuum energy, which are independent of the value of $X_{nl}$~\footnote{ As noted previously, the pseudo-modulus $X_{nl}$ will be stabilized at $X_{nl}=0$ after including the SUSY breaking contributions. So we choose the stabilized value $X_{nl}=0$ for the extrema determined by the condition $H(\tau) = 0$.}. In Fig.\ref{fig1}, we show the energies of the scalar potential for modulus values within the fundamental domain and $X_{nl}=0$, considering various parameter choices.
In the limit $f_0 \to 0$, this scenario reduces to the one involving only a single modulus field $\tau$, whose vacuum is generally not of Minkowski type and the corresponding vacuum energy can possibly be negative. Therefore, for tiny values of $f_0$, the vacuum energy of the global minimum can become negative. Numerical results on the global minima of some benchmark points with non-vanishing vacuum energy are shown in Table~\ref{Minimum:potential}.

\begin{table}[htbp]
		\centering
		\vspace{0.5 cm}
		\begin{tabular}{|c|c|c|c|c|c|c|c|}
			\hline
			$m$ &$n$ & $f_0/M_{\mathrm{Pl}}^{2}$ & $c_1/M_{\mathrm{Pl}}$ & $X_{nl}/M_{\mathrm{Pl}}$& $\tau$&$ P(j(\tau))$&$V_{min}/M_{\mathrm{Pl}}^{4}$ \\\hline
            $0$ & $0$ & $0.1$ & $0.5$ &$  -0.6318 $&
	 		$0.0021+1.002i  $&   $1$&$-0.0525$ \\\hline
			$1$ & $1$ & $10^{-4}$ & $0.1$ &$  0.9994 $&
			$0.19337+0.98275i  $&   $1$&$-1.25$ \\\hline
			$2$ & $0$ & $10^{-7}$ & $-0.01$ &$  -0.9994 $&
			$0.4892+0.8733i  $&   $1$ &$-4.466\times10^{-5}$ \\\hline
			$1$ & $2$ & $1.128\times10^{-6}$ & $-0.01$ &$  -0.3322 $&
			$-0.1910+0.9830i  $&   $1$ &$-7.282\times10^{-5}$ \\\hline
			$2$ & $1$ & $2.974\times10^{-7}$ & $-0.01$ &$  0.2920 $&
			$0.2865+0.9612i  $&   $1$ &$-7.434\times10^{-4}$ \\\hline
            $1$ & $3$ & $2.337\times10^{-6}$ & $-0.01$ &$  -0.4126 $&
			$0.1588+0.9911i  $&   $1$ &$-7.103\times10^{-4}$ \\\hline		
		\end{tabular}       
     \caption{The global minima and the corresponding vacuum energies for some benchmark parameter points in the case $k_X=0$, $f_0\neq 0$ and $c_1\neq 0$. }
     \label{Minimum:potential}
	\end{table}

 From a physical perspective, we are particularly interested in the local minimum with an \textit{intermediate/small} value of $\widetilde{X}$, rather than a global minimum situated at $\widetilde{X} \sim M_{\text{Pl}}$ or $\widetilde{X} \sim  -M_{\text{Pl}}$. For typical benchmark parameter choices, an estimation of the lifetime for the local minimum with an \textit{intermediate/small} value of $\widetilde{X}$ suggests that such a local minimum can be metastable, with a lifetime exceeding the age of the universe. Therefore, we will concentrate on the global (or long-lived meta-stable) minimum with $|X_{nl}|\ll M_{Pl}$ in our subsequent discussions.

 The lifetime of a vacuum can be evaluated semi-classically through a rate of bubble nucleation in unit volume and unit time~\cite{Coleman,Coleman1}. The decay probability, which sets the decay rate, is given by $\Gamma/V=A\exp(-B)$, where $B$ is the action of a Euclidean $O(4)$-symmetric bounce solution, which smoothly interpolates between a true-vacuum configuration at its center
and the false vacuum whose decay is being described. The prefactor $A$ depends on the functional determinants and represents the loop correction of the fluctuations of the fields.  The prefactor $A$ can be approximately given by $A\sim v^4$, where $v$ is some characteristic mass scale of the theory under study. We adopt a conservative choice $v\sim {\cal O}(M_{Pl})$ for our estimation of lifetime.

The generalization to cases involving multiple fields entails determining the bounce configuration by solving a set of coupled ordinary differential equations, subject to the appropriate boundary conditions. For general multi-dimensional field spaces, there will be many bounce solutions with
different actions, and the tunneling rate will be dominated by the solution with the lowest
action. The field profile, which solves the Euclidean equations of motion, traverses a one-dimensional
path in the field space. If one chooses a particular small subset of path in the field space and treats it as a one-field potential, solving the field equations for the one-dimensional potential along this path can give an estimation on the action, which can be close to the actual value of the action~\cite{1702-00356}. 
As the tunneling is expected intuitively to happen through the lowest possible barrier and shortest possible path in the field space, the one with the shortest possible path length which is a straight line between the two minima (or the path with the lowest possible barrier in which a broken line goes through the lowest ridge in the barrier), can act as the one-dimensional approximations of potential along such a path parameterized by the path length in the field space.

 We estimate the decay lifetime with the bounce configuration via the two paths connecting the true and false vacuum for some benchmark parameter choices that can lead to false vacuums at \textit{intermediate/small} value of $\widetilde{X}$. We estimate the one-dimensional action for each case with the approximation of a triangular potential, whose the bounce action is given approximately~\cite{Duncan} by
 \beqa
 B\approx \f{32\pi^2}{3}\f{(\Delta V_+)^2 (\Delta \phi)^4}{\epsilon^3}~.
 \label{bounce:action}
 \eeqa
in the situation where the difference of potential energy between the true and false vacuum $\epsilon$ is very small. Here $\Delta V_\pm=V_{Top}-V_{\pm}$ denotes the height for false (and true) vacuum with respect to the top of the barrier and $\Delta\phi=(\phi_{-}-\phi_+)$ denotes the width of the barrier.

 For a characteristic benchmark point, such as $f_0=10^{-7} M^2_{Pl}$ and $c_1=-0.01 M_{Pl}$ with $m=1,n=1$ and $P(j(\tau))=(j(\tau)-j(\tau_0))^2$ for $\tau_0=1.001\tm e^{i\f{2\pi}{3}(1-\f{1}{36})}$, we find that the false vacuum lies at $(X_{nl},\tau)=(0,1.001\tm e^{i\f{2\pi}{3}\f{35}{36}})$ with vanishing vacuum energy while the true minimum vacuum lies at $(X_{nl},\tau)=(-0.99999 M_{Pl},-0.42825+0.91479i)$ with $V_{min}=-2.49896 M_{Pl}^4$. The peak of the lowest barrier sits at the parameter point $(X_{nl},\tau)=(-0.46285 M_{Pl},-0.40398+0.92017 i)$ with $V_{L-peak}=577.87447 M_{Pl}^4$. We find that the bounce action of the broken line path through the
lowest possible barrier leads to smaller action. It can be calculated with~(\ref{bounce:action}) that the corresponding bounce action $B\approx 2.28\tm 10^5$. Therefore,  the lifetime of the meta-stable vacuum $\tau=(\Ga/V)^{-1/4}$ is much larger than the universe. In fact, using the space-time volume of our past lightcone~\cite{Past:LC}, $(V T)_{LC}= 0.15/H_0^4=3.4 \times 10^{166} {\rm  GeV}^{-4}$ with the Hubble constant $H_0 = 1.44 \times 10^{-42} {\rm GeV}$, the probability that we should have seen a bubble by now is $[(VT)_{LC}] (\Ga/V) \sim 10^{-9670} $, which is exponentially small.



\eit

\subsection{The case with $k_X=12$}

\begin{figure}[hbt]
\begin{center}
\includegraphics[width=2.9 in]{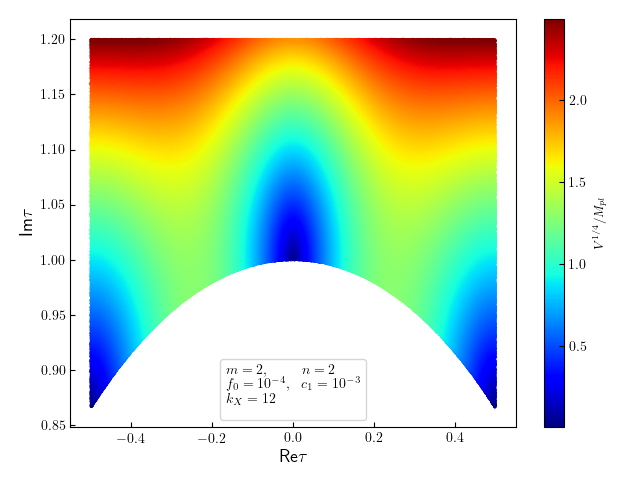}
\includegraphics[width=2.9 in]{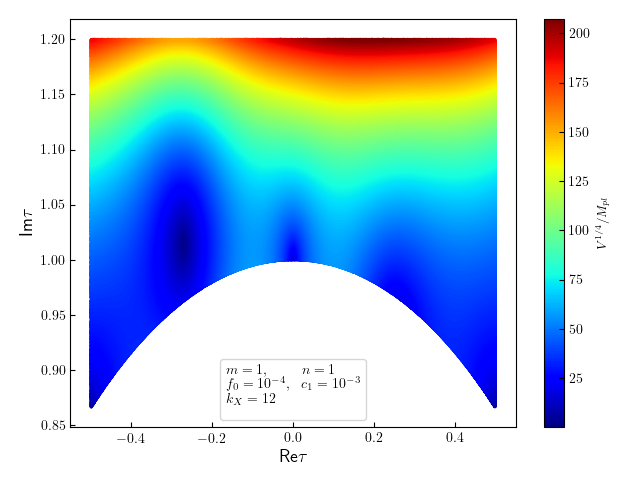}\\
\vspace{-.5cm}\end{center}
\caption{The values of $V(X_{nl},\bar{X}_{nl},\tau,{\tau}^*)$ within the fundamental domain in the scenario with $k_X=12$ and $X_{nl}=0$ for various parameter choices when $f_0\neq 0$ and $f_0\gg c_1^2$. We adopt $\mathcal P(j)=1$ for the left panel and  $\mathcal P(j(\tau))=(j(\tau)-j(\tau_0))^2$ with $\tau_0=1.05\times e^{i\f{\pi}{2}(1+\f{1}{6})}$ for the right panel.}
\label{fig2}
\end{figure}

When $k_X = 12$, interference terms in the scalar potential obscure the analysis of the extrema conditions. Therefore, we leave the most general discussions on this case in our subsequent studies. An important observation is that an $|H(\tau)|^2$ factor (for $H(\tau)\neq 0$) can be extracted from the scalar potential. According to the discussions in the case $k_X=0$, for any value of $X_{nl}$, in addition to the fixed points $\tau=i,\omega$ for most choices of $m,n$, non-fixed point values of modulus VEVs satisfying $|H(\tau)|^2=0$ are local extrema of the scalar potential with vanishing vacuum energy when $P(j(\tau))$ takes the form~(\ref{PjForm:first}). Other local minima of the scalar potential can be determined numerically from the extrema conditions with respect to $\tau$, $X_{nl}$ and the corresponding Hessian matrix. After specifying numerically all other extrema of the scalar potential, we can determined its global minimum or meta-stable minimum by evaluating the corresponding vacuum energies of various local minima (and the lifetime of the meta-stable minimum). We show in Fig.\ref{fig2} the vacuum energy of the scalar potential for typical benchmark parameter choices with modulus values taken in the fundamental domain in the scenario with $k_X=12$.

In our discussion, we assume the existence of modular symmetry in the SUSY breaking sector. The modular symmetry can be broken by SUSY breaking effects, depending on the transformation property of SUSY breaking source under the modular symmetry, as discussed in~\cite{Kikuchi:2022pkd}. For example, in the moduli mediated SUSY breaking scenario~\cite{Kikuchi:2022pkd}, soft SUSY breaking scalar masses and trilinear terms are invariant under the modular symmetry when the F-term is regarded as a spurion with the modular weight $-2$. In our case, non-vanishing F-term VEV of $X_{nl}$ when $k_X\neq 0$ can possibly act as the SUSY breaking source and also break the modular symmetry, whose transformation should be compatible with the modular invariance of $K_{X_{nl}^{\da} X_{nl}}F_{X_{nl}}$, possibly leading to a soft SUSY breaking spectrum that is non-invariant under modular symmetry. The form of the low energy soft SUSY breaking spectrum depends on the concrete mediation mechanism of SUSY breaking that mediates the SUSY breaking effect from the hidden sector to the visible sector.

It should be noted that the modulus stabilization mechanism with nilpotent $\textbf{X}_{nl}$ superfield can be fairly generic and ubiquitous, which is independent of the following discussions on the regeneration of small $\mu_{eff}$ term following modulus stabilization.

\section{\label{sec-5} Regeneration of naturally small $\mu_{eff}$ following modulus stabilization}
In our proposal,  the explicit $\mu$ term in the superpotential is prohibited by modular symmetry. The term involving the $H_uH_d$ bilinear and proper modular form will regenerate an effective $\mu_{eff}$ term after modular symmetry breaking by the VEV of modulus field. It should be noted that the regenerated $\mu_{eff}$ scale is generally not much smaller than $\Lambda_{SUSY}$~\footnote{Strictly speaking, the regenerated $\mu_{eff}$ should lie generically at the scale
\beqa \mu_{eff}\sim \Lambda_{SUSY}F\(\langle\tau\rangle\)=\Lambda_{SUSY}F\(\f{\Lambda_{\not{m}}}{\Lambda_0}\)~, \eeqa where  $\Lambda_{\not{m}}$ is the modular symmetry breaking scale characterized by the modulus VEV $\langle\tau\rangle$, $\Lambda_0$ is the scale appeared in the modulus chiral superfield $\sigma=\Lambda_0\tau$ and $F(\tau)$ is some function approaching the involved modular forms. When $\Lambda_{SUSY}\sim \Lambda_0$ and $F(\tau)\sim \tau$, $\mu_{eff}\sim \Lambda_{\not{m}}$.} unless the stabilized modulus VEV accidentally take some specific value that leads to a large suppression factor. We know that naturalness requires $\mu_{eff}$ to lie at the electroweak scale or soft SUSY breaking scale. So, "generic" values of modulus VEVs from modulus stabilization mechanism that lead to the fully breaking of modular symmetry will in general not lead to a genuine solution of SUSY $\mu$ problem. Therefore, one can consider some "non-generic" values of modulus VEVs that do not break the modular symmetry completely or values of VEVs in the vicinity of such symmetric fixed points.

It is known that the fixed points $\tau_{FP}=i,\omega,i\infty$ for modular theories based on the $\Ga_N$ invariance can preserve $Z_2^S,Z_3^{ST},Z_N^T$ residual symmetries, respectively~\cite{Novichkov:2021evw}. As noted previously,  we can infer from the modular invariance of scalar potential that the finite fixed points should in general be the local extrema of the scalar potential, as they act non-trivially (with residual symmetries) on the derivatives of the potential and must vanish. Although they may not be the global minima, experiences from modulus stabilization in string theory indicate that the minima
should always lie at/near $\tau_{FP}=i,\omega$ for the tree-level scalar potential, as the scalar potential generally diverges for $\tau\ra i\infty$.

It was found in~\cite{Novichkov:2021evw,Feruglio:2021dte} that the hierarchies among the charged leptons and quarks mass can follow solely from the properties of the modular forms (strictly speaking, the representation of the mass term bilinear with respect to the residual group) in the vicinity of symmetric points, avoiding the fine-tuning of the constant parameters (or without the need to introduce extra weighton-like scalar fields). Similar ideas, augmented with additional suppression powers of $q^{1/24}$ (or $(2\Im\tau)^{-1}$), can be used to generate the hierarchy between $\Lambda_{SUSY}$ and $\mu_{eff}$.


Besides, although the scalar potential for modulus field $\tau$ in general diverges when $\tau\ra i\infty$, the presence of $X_{nl}$
contributions can alter the scalar potential, making it possible to stabilize $\tau$ at an intermediate value. Consequently, for an appropriate choice of $k_3$, the effective $\mu_{eff}$ parameter can naturally be much smaller than the SUSY scale $\Lambda_{SUSY}$, due to a large suppression factor originating from both powers of $q^{1/24}$ [or $(2\Im\tau)^{-1}$] and typical suppression behavior of modular forms away from the fixed point $i\infty$ (at $\tau_0$ with medium $\Im\tau_0$).

To regenerate the effective $\mu$-term with a stabilized modulus field, we need to construct a modular weight $-3$ bilinear for $H_u$ and $H_d$ using proper modular forms. For a modular invariant bilinear of the form $\psi_i^c M_{ij}(\tau)\psi_j$, where $M_{ij}(\tau)$ is a modular form of level $N$ and weight $K$, the zero entries of $M_{ij}(\tau)$ at a symmetric point in the mass matrices will generically become non-zero when $\tau$ deviates slightly from the symmetric point. The magnitudes of these residual symmetry breaking entries are controlled by the size of the departure $\epsilon$ from $\tau_{sym}$ and the field transformation properties under the residual symmetry group, which may depend on the modular weights~\cite{Petcov:2023vws,Novichkov:2021evw,Petcov:2022fjf}. That is, the entries of the mass matrices in the vicinity of the symmetric point are subjected to corrections of ${\cal O}(\epsilon^l)$, where the powers $l$ corresponding to the representations of the residual symmetry group
and depends only on how the representations of the fermion fields in the mass term bilinear decompose under the considered residual symmetry group. The degree of suppression, given by the integer $l$, can take values
$l = 0,1,\cdots,N-1$ in the case of $\tau_{sym} = i\infty$, $l = 0,1,2$ in the case of $\tau_{sym} = \omega$ and $l=0,1$ in the case of $\tau_{sym} = i$. In~\cite{Novichkov:2021evw}, the authors systematically classify the residual symmetry representations for different choices of $\Ga_N$ representations of matter fields.

Such a formalism can also be applied to the regeneration of small $\mu_{eff}$ for the $H_u H_d$ bilinear
to achieve a desired suppression factor ${\cal O}(\epsilon^l)$ of the deviation parameter $\epsilon$ (from a given symmetric point). In our case, we need to specify the representations of the Higgs superfields $H_u,H_d$ in MSSM under the residual symmetry group $\mathbb{Z}^T_N$ (with $N=3$) for $\tau_{sym}=i\infty$ and the residual group $\mathbb{Z}^{ST}_3$ for $\tau_{sym}=\omega$. Obviously, $H_u,H_d$ should always be singlets under the modular flavor symmetry $\Ga_N$. They can only be assigned some singlet representations of $\Ga_N$ with non-trivial modular weights. The decompositions of the representation $({\bf r}; k)$ multiplet of $\Ga_N^\pr$ under the residual symmetry can be found in the appendix of~\cite{Novichkov:2021evw}. So, the desired suppression factor for the regenerated $\mu_{eff}$ can be directly obtained with proper choice of singlet representations and modular weights.

We should note that $\mu_{eff}\ll \Lambda_{SUSY}$ can also be naturally realized through a weighton-like mechanism~\cite{King:2020qaj} for $H_u H_d$ bilinear. Specifically, we can introduce a single weighton-like superfield, which is a singlet under both the Standard Model and modular flavor symmetry, with a modular weight $k_\phi=1$. By appropriately choosing the modular weights and representations under the finite modular group $\Ga_N$ for $H_u$ and $H_d$, we can ensure that only the combination
\beqa
W\supseteq \f{Y_{\mathbf{r}}^{(k_\mu)}}{{\[\eta(\tau)\]^{2k_3}}} \(\frac{\phi}{M_F}\)^k H_u H_d~,
\eeqa
is allowed by modular invariance, where $k = k_\mu-k_3 - k_{H_u} - k_{H_d}+3$ is a positive integer, $2k_3\in 24 \mathbb{Z}+ 6$ and $M_F$ is a typical cutoff scale. When the weighton field $\phi$ acquires a VEV through the relevant superpotential, the electroweak-scale $\mu_{\text{eff}}$ can be naturally regenerated with a large suppression factor of order $\tilde{\phi}^k$, where $\tilde{\phi} \equiv \langle \phi \rangle / M_F \ll 1$. For example, the form of superpotential for weigthon superfield can be chosen to be~\cite{King:2020qaj}
 \beqa
 W\supseteq X(Y_{\bf 1}^{4}\f{\phi^4}{M_F^2}-M^2)~,
 \eeqa
with $M$ a dimensional mass parameter and $X$ an $A_4$ singlet. The F-flat condition $F_X^*=-W_X=0$ drives the weighton VEV $\langle \phi\rangle\sim (M M_F)^{1/2}$, which leads to the suppression factor $\frac{\phi}{M_F}\sim (\f{M}{M_F})^{1/2}\ll1$ for $M\ll M_F$.

It should be noted that the $H_u$ and $H_d$ superfields, which transform non-trivially under modular symmetry, can also break the modular invariance after electroweak symmetry breaking.
As the VEVs of $H_u$ and $H_d$ are much smaller than the typical scale of modulus stabilization, the extreme conditions for $\tau$ and $X_{nl}$ will receive negligible corrections from $H_u$ and $H_d$ related terms in the limits $v_u,v_d\ra 0$.
Electroweak symmetry breaking, triggered by the VEVs of $H_u$ and $H_d$, are also affected by the modular stabilization mechanism. However, additional contributions to the electroweak symmetry breaking sector from supergravity vanish in the $M_{Pl}\ra \infty$ limit, leading to negligible contributions to the low energy MSSM scalar potential for $H_u$ and $H_d$, consequently negligible shift of the VEVs for $H_u$ and $H_d$.

\subsection{Natural $\mu_{eff}$ with modulus stabilization near the fixed point $\omega\equiv e^{i2\pi/3}$}

In the vicinity of $\omega$, the degree of suppression for a bilinear is proven~\cite{Novichkov:2021evw} to be given by ${\cal O}(\epsilon^l)$ with $l$ taking values $l=0,1,2$. In this study, we choose to directly calculate the suppression factor and its coefficient for the bilinear $H_u H_d$ using explicit model construction involving appropriate modular forms. It should be emphasized again that the exponentiation power of deviation parameter in the suppression factor depends only on the $\mathbb{Z}^{ST}_3$ (or $\mathbb{Z}^T_3$ for modular VEVs near $i\infty$) charge of $H_u,H_d$ instead of the explicit form of modular forms.

For modular $A_4$ symmetry, at the fixed point $\omega$, the following modular forms satisfy
\beqa
Y_{\bf 1}^{(4)}(\omega)=Y_{\bf 1}^{(8)}(\omega)=Y_{\bf 1^\pr}^{(8)}(\omega)=0~.
\eeqa
 The small deviation of modulus VEV away from $\omega$, induced by the modulus stabilization mechanism, can act as the desired suppression parameter for $\mu_{eff}$.

Expressions of modular forms with larger modular weights can be constructed from $Y_{\bf 3}^{(2)}\equiv (Y_1,Y_2,Y_3)$, for example,
\beqa
Y_{\bf 1^\pr}^{(8)}&=&(Y_1^2+2Y_2Y_3)(Y_3^2+2Y_1Y_2)~,~~~Y_{\bf 1}^{(8)}=(Y_1^2+2Y_2Y_3)^2~.
\eeqa
From the expansion~\cite{Petcov:2022fjf} of modular functions at the point $z_0=\omega+\epsilon$ (with $\epsilon$ depending on the choices of $f_0$ such that $|\epsilon|\ll 1$)
\beqa
\f{Y_2(z_0)}{Y_1(z_0)}&\simeq& \omega\(1+2.235i\epsilon-2.95717\epsilon^2+\cdots\)~,\nn\\
\f{Y_3(z_0)}{Y_1(z_0)}&\simeq& -\f{1}{2}\omega^2\(1+4.47 i\epsilon-10.9096\epsilon^2+\cdots\),
\eeqa
we can obtain the expansion for $Y_{\bf 1}^{(8)}(z_0)$ etc as
\beqa
Y_{\bf 1}^{(8)}(\omega+\epsilon)&\simeq& -36.4138 \epsilon^2-259.129 i \epsilon^3+\cdots~,
\eeqa
with $Y_1(\omega)\approx 0.948674$.

The bilinear term for $H_u H_d$, given in $(\ref{mu:superpotential})$, takes the form
\beqa
W\supseteq \al \Lambda_{SUSY}\f{Y_{\bf r}^{(k_r)}(\tau)}{\[\eta(\tau)\]^{2k_3}} H_u  H_d~,
\label{bilinear}
\eeqa
where $k_{H_u}+k_{H_d}=k_r-k_3+3$, $2k_3\in 24 \mathbb{Z}+ 6$ and ${\bf r}(H_u)\otimes {\bf r}(H_d)\otimes {\bf r}\supset {\bf 1}$. When the modulus field is stabilized close to the point $\omega$, the effective $\mu$-term is given by
\beqa
\mu_{eff}&=& \al \Lambda_{SUSY} \[\eta(\omega+\epsilon)\]^{-2k_3} Y_{\bf r}^{(k_r)}(\omega+\epsilon)(2\Im\tau)^{(k_{H_u}+k_{H_d})/2}~,\nn\\
&=& \al \Lambda_{SUSY}\[\eta(\omega)\]^{-2k_3}Y_{\bf r}^{(k_r)}(\omega+\epsilon)(\sqrt{3})^{(k_r-k_3+3)/2}\nn\\
&\approx&  \al\Lambda_{SUSY} \[0.711506\cdot e^{-0.523599 i}\]^{-k_3/2} Y_{\bf r}^{(k_r)}(\omega+\epsilon)(\sqrt{3})^{k_r+3/2}~.
\eeqa
To achieve a significantly suppressed $\mu_{eff}$, we either need a very large negative value $k_3$ or a tiny $|\epsilon|$. However, calculations show that the suppression from $k_3$ power is not very efficient to obtain a tiny number. Therefore,  we choose to use a small $\epsilon$ to suppress $\mu_{eff}$ with appropriate modular weights and representations for $H_u,H_d$.

For instance, we can choose
\beqa
(r_{H_u},k_{H_u})\sim (~{\bf 1^\pr},~4), ~~(r_{H_d},k_{H_d})\sim (~{\bf 1^{\pr\pr}},~4)~,
\eeqa
and the modular forms $Y_{\bf 1}^{(8)}$ (or, for example, $Y_{\bf 1}^{(4)}$ with suitable $k_{H_u}$ and $k_{H_d}$).
Thus, we have
\beqa
\mu_{eff}&=& \f{\al\Lambda_{SUSY}}{|\bar{N}_{Y_{\bf 1}^{(8)}}|} \[\eta(\omega)\]^{-6} Y_{\bf 1}^{(8)}(\omega+\epsilon)(\sqrt{3})^{(k_{H_u}+k_{H_d})/2}~,\nn\\
&\approx&  -0.927 \tm 10^3 e^{i \f{\pi}{4}}\al \Lambda_{SUSY}\epsilon^2~.
\eeqa
Here $|\bar{N}_{Y_{\bf 1}^{(8)}}|=1.3423$ denotes the global normalization factor~\cite{Petcov:2023fwh} for $Y_{\bf 1}^{(8)}$, which is independent of the modulus VEV and defined by
\beqa
(\overline{{\rm N}}^{(K)}_{\rm Y })^2
=  \int_{D}\int (N^{(K)}_{\rm Y} (\tau,\tau^*))^2\,\frac{dx dy}{y^2}
\equiv \int_{D} \int \sum_i |Y^{(K)}_{{\bf r}i}(\tau)|^2 (2\,y)^{K} \frac{dx\,dy}{y^2}\,,
\label{eq:Global}
\eeqa
with
\beqa
N_Y^{(K)}=\(\sum\limits_{i}\left|Y_i^{(K)}\right|^2 (2 {\rm Im}(\tau))^K\)^{1/2}~,
\eeqa
or
\beqa
\begin{aligned}
(\overline{{\rm N}}^{(K)}_{\rm Y R})^2
& = \lim_{T\to \infty}
\left ( \int_{D_T} \int (N^{(K)}_{\rm Y} (\tau,\tau^*))^2\,\frac{dx dy}{y^2}
- |a_0|^2 {\rm \frac{2^{K}}{K-1}\, T^{K-1} } \right )
\\
& \equiv
\lim_{T \to \infty}
\left (\int_{D_T} \int \sum_i |Y^{(K)}_{{\bf r}i}(\tau)|^2 (2\,y)^{\rm K} \frac{dx\,dy}{y^2}
- |a_0|^2 {\rm \frac{2^{K}}{K-1}\,T^{K-1} }\right )\,,
\label{eq:non-cusp}
\end{aligned}
\eeqa
%
for non-cusp modular forms, where $T$ is a real constant, $D_T$ is the region below the line $y=T$ in the fundamental domain with $T > 1$, and $|a_0|^2$ is the constant $q$-independent term in the $q$-expansion of $\sum_i |Y^{(K)}_{{\bf r}i}(\tau)|^2$. It should be noted that the normalization can not be fixed in the bottom-up approach to modular symmetry.

From the decompositions of $\Ga_N$ representations under the residual symmetry group in appendix of~\cite{Novichkov:2021evw}, it can be verified that the suppression factor of $\epsilon$ for the $H_u(~{\bf 1^\pr},~4)$ and $H_d(~{\bf 1^{\pr\pr}},~4)$ bilinear is given by $({k_{H_u}+k_{H_d}+3})~{\rm modulo}~3=2$, which is consistent with the result from our explicit construction.

If $\Lambda_{SUSY}$ is naively taken to be the Planck scale, a tiny $\epsilon$ with $|\epsilon|\sim 10^{-10}$ or a tiny $\al$ is needed to achieve an electroweak-scale $\mu_{eff}$. For a smaller $\Lambda_{\text{SUSY}}$ scale, a smaller suppression factor is required to achieve an effective $\mu_{\text{eff}}$ at the electroweak scale. For instance, in gravity mediated SUSY breaking mechanism, where the SUSY breaking scale $\Lambda_{SUSY}$ is typically of the order $\sqrt{F}\sim\sqrt{m_{3/2}M_{Pl}}\sim 10^{10}$ GeV for $m_{3/2}\sim {\cal O}({\rm 10^2~GeV})$,
a desired $\mu_{eff}$ at electroweak scale can be regenerated for $|\epsilon|\sim 10^{-5}$ when $\al\sim {\cal O}(1)$. In gauge mediated SUSY breaking mechanism where the SUSY breaking scale is as low as $10^5$ GeV, the required suppression factor $\epsilon$ can be of the order $|\epsilon| \sim 10^{-2}$ for $\alpha \sim \mathcal{O}(1)$.

Note that the smallness of deviation parameter $\epsilon$ (away from $\omega$) can be a consequence of SUSY breaking, which depends on the choice of $f_0$. For example, if the parameter choice with $f_0=0$ leads to the global minimum at $\omega$, SUSY breaking contributions may slightly shift the minimum to some point near $\omega$.

The desired condition $|\epsilon| \ll 1$ can be achieved through typical modulus stabilization mechanisms, such as the one proposed in~\cite{NPP:2201.02020}, which employs large values of $m$ and $n=0$. However, for the modulus stabilization mechanism discussed in the previous section, fine-tuning is required to obtain a minimum at $\omega + \epsilon$ with $|\epsilon| \ll 1$. This is particularly true when we consider the simplest realization ${\cal P}(j(\tau)) =(j(\tau) - c_p)^2$ with $c_p \ll 1$ for $H(\tau) = 0$. We provide in Table~\ref{omega:suppression} some benchmark points that can achieve small suppression factors using the aforementioned modulus stabilization mechanism. It is worth noting that the fine-tuning needed here may arise from the specific choices of input parameters for our modulus stabilization mechanism. One could explore alternative modulus stabilization mechanisms to achieve a minimum very close to $\omega$ without large fine-tuning. Moreover, such fine-tuning is not necessary when employing a weighton-type mechanism for the regeneration of $\mu_{\text{eff}}$.\\

\begin{table}[htbp]
		\centering
		\vspace{0.5 cm}
				\begin{tabular}{|c|c|c|c|c|c|}
			\hline
			$m$ &$n$ & $f_0/M_{\mathrm{Pl}}^{2}$ & $c_1/M_{\mathrm{Pl}}$ & $\epsilon/10^{-5}$&$ P(j(\tau))$ \\\hline
			$2$ & $2$ & $0.2$ & $0.1$ &$  (1.659+0.958i)$&   $(j-1.765\times 10^{-9})^2$ \\\hline
			$0$ & $3$ & $0.2$ & $0.1$ &$  (5.441+3.141i)$&   $(j-2.861\times 10^{-9})^2$  \\\hline
			$1$ & $2$ & $0.2$ & $0.1$ &$  (8.162+4.172i)$&   $(j-3.973\times 10^{-8})^2$  \\\hline
			$0$ & $2$ & $0.2$ & $0.1$ &$  (6.802+3.927i)$&   $(j-280.469)(j-1.008\times 10^{-8})^2$  \\\hline
			$2$ & $3$ & $0.2$ & $0.1$ &$ i $&   $(j-1.398\times 10^{-9})^2$ \\\hline
		\end{tabular}
\caption{Parameter choices for the scenario with $k_X=0$, which can stabilize the modulus at the point $\omega+\epsilon$ (for $|\epsilon|\sim 10^{-5}|$) with $X_{nl}=0$.}
\label{omega:suppression}
\end{table}

\subsection{Natural $\mu_{eff}$ with modulus stabilization at medium $\Im\tau$ }

When the modulus field is stabilized near $i\infty$, similar to the case where modulus VEV lies in the vicinity of $\omega$, the exponentiation
power $l$ of deviation parameter $\epsilon$ in the suppression factor for the $H_u H_d$ bilinear can also be proven~\cite{Novichkov:2021evw} to depend only on the $\mathbb{Z}^T_3$ charge of $H_u, H_d$, rather than the explicit form of modular forms. Since the coefficient of $\epsilon^l$ cannot be simply determined and may involve additional suppression factors, we opt to directly calculate the suppression factor and the coefficient for the bilinear $H_uH_d$  using an explicit model construction.

Although the scalar potential for the modulus generally diverges as $\tau\ra i\infty$, the introduction of nilpotent Goldstino $\textbf{X}_{nl}$ superfield or multiple moduli/matter fields may alter the shape of the scalar potential at intermediate values of $\tau$, making it possible to stabilize the modulus at $\tau_0$ with an intermediate value of $\Im\tau_0$ (for example, $\Im\tau_0\sim 5-10$). Detailed discussions on such a possibility will be given in our subsequent studies~\footnote{For example, we can choose the form ${\cal P}(j(\tau))=(j(\tau)-j(i \Im\tau_0 ))^2$ (or other polynomials that contain such a factor) so as that $i\Im\tau_0$ is a solution of ${\cal P}(j(\tau))={\cal P}^\pr(j(\tau))=0$, consequently being an extrema of the scalar potential. It is also possible to change the shape of scalar potential near $i\infty$ by introducing non-minimal gravitational couplings of modulus fields to Ricci scalar~\cite{Nonmimimal:GR}. After transforming back from the Jordan frame to Einstein frame, the modulus field can be stabilized at intermediate values of $\tau$ more easily with the transformed scalar potential.}.  In this work, we just concentrate on the implications of modulus stabilization at an intermediate value of $\Im\tau$ on the SUSY $\mu$-problem. Note that the stabilized VEV $\tau_0$ (with an intermediate value $\Im\tau_0$) can be regarded as lying close to the $i\infty$ fixed point.

To obtain the suppressed effective $\mu$-parameter, in addition to possible suppression factor from powers of $q^{1/24}$ (from $\eta(\tau)$) or $(2\Im \tau)^{-1}$ (from normalization factor), we would also like to choose the modular function that vanishes at $\Im \tau =i \infty$. For example, the modular forms satisfy
\beqa
Y_{\bf 1^\pr}^{(4)}(i\infty)=Y_{\bf 1^\pr}^{(8)}(i\infty)=Y_{\bf 1^{\pr\pr}}^{(8)}(i\infty)=0~,
\eeqa
as well as the vanishing singlet components from their direct products, such as the modular function $Y_{\bf 1}^{(12)}(i\infty)$.

The asymptotic behavior of the previous modular forms near $i\infty$ within the fundamental domain can be calculated from~\cite{Petcov:2023vws} (for $z_0$ with $\Im z_0$ sufficiently large)
\beqa
Y_{\bf 1^\pr}^{(4)}(z_0)&=&-12 p \epsilon+96 p^4 \epsilon^4+{\cal O}(\epsilon^7)~,\nn\\
Y_{\bf 1^\pr}^{(8)}(z_0)&=&-6 p \epsilon-1230 p^4 \epsilon^4+{\cal O}(\epsilon^7)~,\nn\\
Y_{\bf 1^{\pr\pr}}^{(8)}(z_0)&=&144 p^2 \epsilon^2-2304 p^5 \epsilon^5+{\cal O}(\epsilon^8)~,
\eeqa
with
\beqa
\epsilon=\exp\[-\f{2\pi}{3}\Im z_0\]~,~~p=\exp\[i\f{2\pi}{3}\Re z_0\],
\eeqa
where $\epsilon\ll1$.

Let us suppose that the modulus field could be stabilized at the imaginary axis (within the fundamental domain) with $z_0=it_0$ for medium large $t_0$, when proper SUSY breaking contributions with non-vanishing $f_0$ are taken into account. Naively, if the desired suppression factor to regenerate $\mu_{eff}$ comes solely from $\epsilon^2$ when $\Lambda_{SUSY}$ is naively taken to be the Planck scale, very large $t_0$ (here $t_0\gtrsim 10$) is needed. Such a large $t_0$ possibly causes difficulties for modulus stabilization realizations, as the scalar potential for modulus field generally diverges for $\tau\ra i\infty$. Therefore, it is desirable to introduce additional suppression factors other than $\epsilon$ so that medium large $t_0$ can lead to desired suppression factors.

 The effective $\mu$-term from the bilinear term for $H_u H_d$ in~(\ref{bilinear}), which satisfies $k_{H_u}+k_{H_d}=k_r-k_3+3$ for $2k_3\in 24 \mathbb{Z}+ 6$, is given by
\beqa
\mu_{eff}&=& \al\Lambda_{SUSY}\[\eta(it_0)\]^{-2k_3} Y_{\bf r}^{(k_r)}(it_0) (2t_0)^{(k_{H_u}+k_{H_d})/2}~,\nn\\
&\approx& \al \Lambda_{SUSY} e^{\f{ k_3 \pi t_0}{6}}(2t_0)^{-\f{k_3}{2}} Y_{\bf r}^{(k_r)}(it_0) (2t_0)^{\f{k_r+3}{2}}~,
\eeqa
where proper normalization factors for $H_u$,$H_d$ from the Kahler potential are taken into account.

For fixed $k_r$, we require that the factor
\beqa
\[e^{\f{\pi t_0}{3}}(2t_0)^{-1}\]^{\f{k_3}{2}} \ll 1,
\eeqa
is tiny to act as an additional suppression factor other than $\epsilon$. As the factor
\beqa
e^{\f{\pi t_0}{3}}(2t_0)^{-1}>1~,
\eeqa
for all $t_0>0$, the value of $k_3$ needs to be negative.

We can choose the following modular weights and representations
\beqa
(r_{H_u},k_{H_u})\sim (~{\bf 1},~16), ~~(r_{H_d},k_{H_d})\sim (~{\bf 1^{\pr}},~16)~,
\eeqa
 and the choice of modular form $Y_{\bf 1^{\pr\pr}}^{(8)}$ (or, for example, $Y_{\bf 1}^{(12)}$, $Y_{\bf 1^\pr}^{(16)}$ etc with proper $k_{H_u}$ and $k_{H_d}$) with $k_3=-21$ to generate a suppressed $\mu_{eff}$ parameter.
In this case, the effective $\mu_{eff}$ parameter is
\beqa
\mu_{eff}&=&\f{\al}{\bar{N}_{Y_{\bf 1^{\pr\pr}}^{(8)}}} \Lambda_{SUSY} \[e^{\f{\pi t_0}{3}}(2t_0)^{-1}\]^{-21/2}144 e^{-\f{4\pi t_0}{3}}(2t_0)^{11/2}~,\nn\\
&=&7.05 {\al}\Lambda_{SUSY}  e^{-29\pi t_0/6}(2t_0)^{16}~,
\eeqa
where ${\bar{N}_{Y_{\bf 1^{\pr\pr}}^{(8)}}}=20.4152$ is the global normalization factor for the modular form $Y_{\bf 1^{\pr\pr}}^{(8)}$.

Calculations show that the value $\mu_{eff}\lesssim 10^{-17} \Lambda_{SUSY}$ can be achieved for $t_0\gtrsim 5.01$, if $\Lambda_{SUSY}$ is naively taken to be the Planck scale. For $\Lambda_{SUSY}$ typically of the order $\sqrt{F}\sim\sqrt{m_{3/2}M_{Pl}}\sim 10^{10}$ GeV, it is sufficient to choose $t_0\gtrsim 3.2 $ to satisfy $\mu_{eff}\lesssim 10^{-8} \Lambda_{SUSY}$. For a smaller $\Lambda_{SUSY}$ scale, a smaller suppression factor is needed to achieve an effective $\mu_{eff}$ at the electroweak scale, requiring a smaller value of $t_0$. The tiny value of $\mu_{eff}$ results from the combined effect of suppression by powers of $q$ (or $(2\Im\tau)^{-1}$) and the asymptotic suppression behavior of typical modular forms away from the fixed point $i\infty$.

 It can also be verified that the suppression power of $\epsilon=\exp[-\f{2\pi}{3}\Im z_0]$ for $H_u(~{\bf 1},~16)$ and $H_d(~{\bf 1^{\pr}},~16)$ bilinear is given by the power factor $l$ of $(\rho_i^c\rho_j)^*=\zeta^l$ for $\zeta=\exp(2\pi i/3)$, which can be read out from~\cite{Novichkov:2021evw} to be $(1\cdot\zeta)^*=\zeta^2$ and predict $l=2$. This power factor for $\epsilon$ is consistent with the result from our explicit construction.

\section{\label{sec:conclusions} Conclusions}

In this study, we propose a novel solution to the SUSY $\mu$-problem within the framework of modular flavor symmetry. The explicit $\mu$-term is prohibited by modular symmetry after assigning non-vanishing modular weights and/or non-trivial singlet representations to the $H_u,H_d$ superfields. The effective $\mu$-term is then regenerated following the stabilization of the modulus field, with the corresponding modulus VEV breaks the modular symmetry partially (at the fixed points) or completely (in the vicinity of the fixed points). Since naturalness requires $\mu_{eff}$ to lie at the electroweak scale or soft SUSY breaking scale, a large suppression factor leading to $\mu_{eff}\ll \Lambda_{SUSY}$ is required when the modulus field is stabilized. Such a large suppression factor can naturally arise from a modulus VEV in the vicinity of certain fixed points or by the suppression factor in weighton-like mechanism.

Ordinary single modulus stabilization mechanism can be affected by SUSY breaking contributions. The Goldstino, which is necessarily present as a consequence of UV SUSY breaking and mixes with gravitino, can be promoted into the constrained nilpotent superfield $\textbf{X}_{nl}$ and provides a way for an arbitrary non-supersymmetric low-energy theory to realize SUSY non-linearly.
 We discuss the stabilization mechanism of a single modulus field in the presence of SUSY breaking contributions, described by the non-linear SUSY realization scheme with a nilpotent Goldstino $\textbf{X}_{nl}$.

  Natural value of $\mu_{eff}$, which is significantly suppressed relative to the SUSY scale, can be achieved from the expansion of typical modular forms using a tiny deviation parameter near the fixed points. This is similar to the mechanism that can naturally generate mass hierarchies for quarks and leptons in the vicinity of certain fixed points. The exponentiation
power of the deviation parameter in the suppression factor, in fact, depends solely on the $\mathbb{Z}_3^{ST}$ charge (for fixed point $\omega$), or $\mathbb{Z}^T_3$ charge (for fixed point $i\infty$) of $H_u$, $H_d$, rather than the explicit form of modular forms.

For a modulus VEV with an intermediate value of $\Im\tau$,
  the significantly suppressed $\mu_{eff}$ results from the combined effects of suppression by powers of $q^{1/24}$ [or $(2\Im\tau)^{-1}$], where the exponentiation power determined by the modular weights of $H_u$ and $H_d$, along with the asymptotic suppression behavior of typical modular forms away from the fixed point $i\infty$, taking the form of appropriate power of the tiny deviation parameter.

  The modulus stabilization mechanism with nilpotent $\textbf{X}_{nl}$ superfield can be fairly generic and ubiquitous. The regeneration of small $\mu_{eff}$ term following modulus stabilization, which just needs a modulus VEV near some fixed points or through a weighton-like mechanism for $H_u H_d$ bilinear, is independent of the concrete modulus stabilization mechanism.

\appendix
\section{Properties of Modular $A_4$}
The finite modular group $A_4\cong \Ga_3$ can generated by
\beqa
S^2=(ST)^3=T^3=1~.
\eeqa
There are 4 inequivalent irreducible representations of $A_4$: three singlets $\mathbf{1}$, $\mathbf{1}'$, $\mathbf{1}''$ and a triplet $\mathbf{3}$.

The triplet representation $\mathbf{3}$ (in the basis where $T$ is diagonal) is given by
\begin{equation}
\label{eq:rep-gen}S=\frac{1}{3}\begin{pmatrix}
-1 ~& 2  ~& 2  \\
2  ~& -1 ~& 2 \\
2  ~& 2  ~& -1
\end{pmatrix}, ~\quad~
T=\begin{pmatrix}
1 ~&~ 0 ~&~ 0 \\
0 ~&~ \omega ~&~ 0 \\
0 ~&~ 0 ~&~ \omega^{2}
\end{pmatrix} \,.
\end{equation}

The decompositions of the direct product of $A_4$ representations are
\begin{eqnarray}
\nonumber&& \mathbf{1}'\otimes\mathbf{1}'=\mathbf{1}'',~~~ \mathbf{1}'\otimes\mathbf{1}''=\mathbf{1},~~~ \mathbf{1}''\otimes\mathbf{1}''=\mathbf{1}'\,,\\
&&\mathbf{3}\otimes \mathbf{3}= \mathbf{1}\oplus \mathbf{1'}\oplus \mathbf{1''}\oplus \mathbf{3}_S\oplus \mathbf{3}_A\,,
\label{A4:33product}
\end{eqnarray}
where $\mathbf{3}_{S(A)}$ denotes the symmetric (antisymmetric) combinations, respectively.

Given two $A_4$ triplet representations $\alpha=(\alpha_1,\alpha_2,\alpha_3)$ and  $\beta=(\beta_1,\beta_2,\beta_3)$, the irreducible representations obtained from their direct products are:
\begin{eqnarray}
\nonumber &&\mathbf{1}=\alpha_1\beta_1+\alpha_2\beta_3+\alpha_3\beta_2\,, \\
\nonumber &&\mathbf{1}'=\alpha_3\beta_3+\alpha_1\beta_2+\alpha_2\beta_1\,, \\
\nonumber &&\mathbf{1}''=\alpha_2\beta_2+\alpha_1\beta_3+\alpha_3\beta_1\,, \\
\nonumber &&\mathbf{3}_S=(
2\alpha_1\beta_1-\alpha_2\beta_3-\alpha_3\beta_2,
2\alpha_3\beta_3-\alpha_1\beta_2-\alpha_2\beta_1,
2\alpha_2\beta_2-\alpha_1\beta_3-\alpha_3\beta_1)\,, \\
\label{eq:decomp-rules} &&\mathbf{3}_A=(
\alpha_2\beta_3-\alpha_3\beta_2,
\alpha_1\beta_2-\alpha_2\beta_1,
\alpha_3\beta_1-\alpha_1\beta_3)\,.
\end{eqnarray}

Modular weight $k=2$ and the representation ${\bf r}={\bf 3}$ are given as
$Y^{(2)}_{\mathbf{3}}=\left(Y_1, Y_2, Y_3\right)^{T}$
 with
\beqa
Y_1(\tau) &=& \f{i}{2\pi}\left[ \frac{\eta'(\tau/3)}{\eta(\tau/3)}  +\frac{\eta'((\tau +1)/3)}{\eta((\tau+1)/3)}
+\frac{\eta'((\tau +2)/3)}{\eta((\tau+2)/3)} - \frac{27\eta'(3\tau)}{\eta(3\tau)}  \right], \nonumber \\
Y_2(\tau) &=& \frac{-i}{\pi}\left[ \frac{\eta'(\tau/3)}{\eta(\tau/3)}  +\omega^2\frac{\eta'((\tau +1)/3)}{\eta((\tau+1)/3)}
+\omega \frac{\eta'((\tau +2)/3)}{\eta((\tau+2)/3)}  \right] ,\nonumber \\
Y_3(\tau) &=& \frac{-i}{\pi}\left[ \frac{\eta'(\tau/3)}{\eta(\tau/3)}  +\omega\frac{\eta'((\tau +1)/3)}{\eta((\tau+1)/3)}
+\omega^2 \frac{\eta'((\tau +2)/3)}{\eta((\tau+2)/3)} \right]\,,
\eeqa
where $\eta(\tau)$ is the Dedekind eta-function,
\begin{equation}
\eta(\tau)=q^{1/24} \prod_{n =1}^\infty (1-q^n), ~~~  q=e^{2\pi i\tau}\,.
\end{equation}

The $q-$expansions of $Y_{1,2,3}(\tau)$ are given as
\beqa
&&Y_1(\tau)=1 + 12q + 36q^2 + 12q^3 + 84q^4 + 72q^5 +\dots\,,\nn\\
&&Y_2(\tau)=-6q^{1/3}(1 + 7q + 8q^2 + 18q^3 + 14q^4 +\dots)\,,\nn\\
&&Y_3(\tau)=-18q^{2/3}(1 + 2q + 5q^2 + 4q^3 + 8q^4 +\dots)\,.
\eeqa
\section{Eisenstein Series\label{appendix:B}}
The Eisenstein series $G_k(\tau)$, defined by
\beqa
G_k(\tau)=\sum\limits_{m,n\in Z}^{\pr}\f{1}{\(m+n\tau\)^{k}}=\zeta(k)\sum\limits_{\begin{subarray} a c,d\in {Z}; \\gcd(c,d)=1 \end{subarray}}\f{1}{(c\tau+d)^k}~,
\eeqa
is a modular form of weight $k$ for $\tau\in {\cal H}$ and has the following Fourier decomposition
\beqa
 G_k(\tau)=2\zeta(k)+2\f{(2\pi i)^k}{\Ga(k)}\sum\limits_{N=1}^\infty\sigma_{k-1}(N)q^N~,
\eeqa
with $\sigma_{\al}(N)$ the sums of divisors functions, defined for any $\al\in \mathcal{C}$ by
\beqa
\sigma_{\al}(N)=\sum\limits_{l|N}l^\al~.
\eeqa
It can be proven that $G_4(\rho)=G_6(i)=0$.

Factoring out $2\zeta(k)$, and using the relation between Bernoulli numbers $B_k$ and $\zeta(k)$ for even $k$, the normalized Eisenstein series $E_k$ can be defined by
\beqa
E_k(\tau)=\f{G_k(\tau)}{2\zeta(k)}=1-\f{2k}{B_k}\sum\limits_{N=1}^\infty\sigma_{k-1}(N)q^N~,
\eeqa
with some of the $\nu_k\equiv-{2k}/{B_k}$ values given by
\beqa
\nu_4 = 240,~ \nu_6 = 504,~ \nu_8 =480.
\eeqa
The quasi-modular transformation property of $E_2(\tau)$
\beqa
E_2(\ga \tau)&=&(c\tau+d)^{2}E_2(\tau)+\f{12}{2\pi i}c(c\tau+d)~,
\eeqa
 allows one to evaluate $E_2(\tau)$ at the points $\tau=i$ and $\tau = \omega$.

The values of $E_k(\tau)$ are listed below~\cite{DHoker:2022dxx}
\beqa
E_2(i)&=&\f{3}{\pi},~~E_2(\omega)=\f{2\sqrt{3}}{\pi},~~E_4(i)=\f{48[\Ga(\f{5}{4})]^4}{\pi^2[\Ga(\f{3}{4})]^4},~\nn\\
E_4(\omega)&=& 0,~~E_6(i)=0,~~E_6(\omega)=\f{729}{2\pi^3}\f{[\Ga(\f{4}{3})]^6}{[\Ga(\f{5}{6})]^6}.
\eeqa

The second derivative of $j(\tau)$ are given by
\beqa
j^{\pr\pr}(\tau)=\f{4\pi^2}{[\eta(\tau)]^{24}}\left\{\f{1}{6}E_2(\tau)[E_4(\tau)]^2E_6(\tau)-\f{2}{3}E_4(\tau)[E_6(\tau)]^2-\f{1}{2}[E_4(\tau)]^4\right\}~,
\eeqa
from Ramanujan's formulas for derivatives of Eisenstein series, which gives
\beqa
j^{\pr\pr}(\omega)=0~,~~j^{\pr\pr}(i)=-\f{2\pi^2}{[\eta(i)]^{24}}[E_4(i)]^4~.
\eeqa

\section{Solutions of $H^\pr(\tau)=0$\label{appendix:C}}

Such a constraint reduces to
\begin{equation}
H'(\tau) = \left\{j(\tau)^{n/3}[j(\tau)-1728]^{m/2} \mathcal{P}(j(\tau))\right\}' = 0,
\end{equation}
or equivalently,
\begin{equation}
H(\tau) j'(\tau) \left[ \frac{n}{3} \frac{1}{j(\tau)} + \frac{m}{2} \frac{1}{j(\tau)-1728} + \frac{1}{\mathcal{P}(j(\tau))} \frac{\partial \mathcal{P}(j(\tau))}{\partial j(\tau)} \right] = 0.
\label{Hprime:zero}
\end{equation}

Note that the expression
\begin{equation}
H(\tau)\left[ \frac{n}{3} \frac{1}{j(\tau)} + \frac{m}{2} \frac{1}{j(\tau)-1728} + \frac{1}{\mathcal{P}(j(\tau))} \frac{\partial \mathcal{P}(j(\tau))}{\partial j(\tau)} \right],
\label{constraint}
\end{equation}
can be singular at certain solutions of $H(\tau)=0$. Such singular behavior would be eliminated by the multiplication of $j'(\tau)$ to render $H^\pr(\tau)$ finite at such points, which can be seen in (\ref{Hprime:iomega}).

Firstly, we would like to discuss the case $H(\tau)\neq 0$. From equation (\ref{constraint}), it is obvious that either
\beqa
 \left[ \frac{n}{3} \frac{1}{j(\tau)} + \frac{m}{2} \frac{1}{j(\tau)-1728} + \frac{1}{\mathcal{P}(j(\tau))} \frac{\partial \mathcal{P}(j(\tau))}{\partial j(\tau)} \right]=0~,
\label{Hprime:equation}
\eeqa
or $j'(\tau)=0$.

When $H(\tau)\neq 0$, for non-negative integers $m$ and $n$ (by the definition of $H(\tau)$), we have the following discussions for equation (\ref{Hprime:equation}) after eliminating the $H(\tau)$ factor:
\begin{itemize}
    \item $m + n > 0$:
    \begin{itemize}
        \item When $n \neq 0$ and $m = 0$, condition (\ref{Hprime:equation}) reduces to
        \begin{equation}
        \frac{d \mathcal{P}(j(\tau))}{\mathcal{P}(j(\tau))} = -\frac{n}{3} \frac{d j(\tau)}{j(\tau)},
        \end{equation}
        which gives
        \begin{equation}
        \mathcal{P}(j(\tau)) = C_0 [j(\tau)]^{-n/3}.
        \end{equation}
        So, $\mathcal{P}(j(\tau))$ is not a polynomial of $j(\tau)$ unless $n \in 3\mathbb{Z}$ is negative, contradicting the assumption that $n$ is non-negative integers.

        \item When $m \neq 0$ and $n = 0$, it can also be deduced that $\mathcal{P}(j(\tau))$ is a polynomial of $j(\tau)$ unless $m \in 2\mathbb{Z}$ is negative, contradicting the assumption that $m$ is non-negative integers. 

        \item For $m \neq 0$ and $n \neq 0$, similar deduction gives
        \begin{equation}
        \mathcal{P}(j(\tau)) = C_0' [j(\tau)]^{-n/3} [j(\tau)-1728]^{-m/2}.
        \end{equation}
        So, $\mathcal{P}(j(\tau))$ is not a polynomial of $j(\tau)$ unless both $m \in 2\mathbb{Z}$ and $n \in 3\mathbb{Z}$ are negative. As $m, n$ are defined to be non-negative integers, this condition requires $m = n = 0$, contradicting the assumptions that $m,n\neq 0$.
    \end{itemize}
     Therefore, $m+n>0$ can not hold.
    \item $m = 0$ and $n = 0$:

    With the constraints $m = n = 0$, the condition (\ref{constraint}) reduces to
    \begin{equation}
    \frac{\partial \mathcal{P}(j(\tau))}{\partial j(\tau)} = 0,
    \end{equation}
    which gives
    \begin{equation}
    \mathcal{P}(j(\tau)) = \text{Constant}.
    \end{equation}
 Since $H(\tau) \neq 0$, this constant must therefore be non-zero.
\end{itemize}

When $H(\tau)\neq 0$, for the condition $j^\pr(\tau)=0$, the derivative of the Klein $j$-function is given by
\beqa
\f{\pa}{\pa \tau} j(\tau)=-2\pi i\f{{E_6(\tau)}{E_4^2(\tau)}}{[\eta(\tau)]^{24}}=0~.
\eeqa
It can checked that its solutions are $\omega$ and $i$, contradicting the assumption $H(\tau)\neq 0$ with $m=n=0$.  Therefore, the condition $H'(\tau) = 0$ with $H(\tau) \neq 0$ implies that $H(\tau)$ is a non-zero constant.

The solutions of $H'(\tau) = 0$ for the case $H(\tau) = 0$ are addressed in the paragraphs containing equations~(\ref{PjForm:m}), (\ref{PjForm:n}), and (\ref{PjForm:first}).

\begin{acknowledgments}
These authors contributed equally: Hong Jie Fan, Ying Kai Zhang. We are very grateful to the referees for helpful discussions. This work was supported by the National Natural Science Foundation of China under grant numbers Nos.12075213 and 12335005, by the Natural Science Foundation of Henan (Distinguished Young Scholars of Henan Province) under grant number 242300421046.
\end{acknowledgments}

\end{document}